\documentclass[twocolumn,tighten,trackchanges]{aastex631}

\definecolor{amethyst}{rgb}{0.8, 0.0, 0.0}
\definecolor{darkgreen}{rgb}{0,0.35,0}

\definecolor{coral}{rgb}{0,0.75,0}
\definecolor{green}{rgb}{0., 0.75, 0.}
\definecolor{black}{rgb}{0., 0., 0.}
\newcommand{\rev}[1]{\textcolor{black}{#1}}

\definecolor{vale}{RGB}{154,0,154}

\definecolor{boh}{RGB}{10,180,20}

\definecolor{green}{rgb}{0,1,0}

\definecolor{verde}{RGB}{212, 195, 15}

\newcommand{\NEXUS}{\textsc{NEXUS+}}

\def\hmsun{h^{-1} {\rm{M_\odot}}}

\def\nobhdi{$f_{\rm BH, \, DI}=0$ }

\defcitealias{Parente2023}{P23}

\usepackage{soul}
\usepackage{ulem}
\usepackage{cancel}
\usepackage{graphicx}	
\usepackage{amsmath}	
\usepackage{soul}
\usepackage{color}
\usepackage{natbib}
\usepackage{url}
\usepackage{placeins}
\usepackage{xfrac}
\usepackage{comment}
\usepackage{multirow}
\usepackage{amssymb}
\usepackage{booktabs}
\usepackage{soul}
\usepackage{tabularx}
\usepackage{verbatim}

\DeclareRobustCommand{\VAN}[3]{#2}
\let\VANthebibliography\thebibliography
\def\thebibliography{\DeclareRobustCommand{\VAN}[3]{##3}\VANthebibliography}

\begin{document}

\title{Star Formation and Dust in the Cosmic Web}

\author[0000-0002-9729-3721]{Massimiliano Parente}
\affiliation{SISSA, Via Bonomea 265, I-34136 Trieste, Italy}
\affiliation{INAF, Osservatorio Astronomico di Trieste, via Tiepolo 11, I-34131, Trieste, Italy}

\author[0000-0003-2826-4799]{Cinthia Ragone-Figueroa}
\affiliation{Instituto de Astronom\'ia Te\'orica y Experimental (IATE), Consejo Nacional de Investigaciones Cient\'ificas y T\'ecnicas de la\\ Rep\'ublica Argentina (CONICET), Universidad Nacional de C\'ordoba, Laprida 854, X5000BGR, C\'ordoba, Argentina}
\affiliation{INAF, Osservatorio Astronomico di Trieste, via Tiepolo 11, I-34131, Trieste, Italy}

\author[0000-0002-9596-9812]{Pablo López}
\affiliation{Instituto de Astronom\'ia Te\'orica y Experimental (IATE), Consejo Nacional de Investigaciones Cient\'ificas y T\'ecnicas de la\\ Rep\'ublica Argentina (CONICET), Universidad Nacional de C\'ordoba, Laprida 854, X5000BGR, C\'ordoba, Argentina}
\affiliation{Observatorio Astronómico, Universidad Nacional de Córdoba, Laprida 854, X5000BGR, Córdoba, Argentina}

\author[0000-0003-0477-5412]{Héctor J. Martínez}
\affiliation{Instituto de Astronom\'ia Te\'orica y Experimental (IATE), Consejo Nacional de Investigaciones Cient\'ificas y T\'ecnicas de la\\ Rep\'ublica Argentina (CONICET), Universidad Nacional de C\'ordoba, Laprida 854, X5000BGR, C\'ordoba, Argentina}
\affiliation{Observatorio Astronómico, Universidad Nacional de Córdoba, Laprida 854, X5000BGR, Córdoba, Argentina}

\author[0000-0001-5035-4913]{Andrés N. Ruiz}
\affiliation{Instituto de Astronom\'ia Te\'orica y Experimental (IATE), Consejo Nacional de Investigaciones Cient\'ificas y T\'ecnicas de la\\ Rep\'ublica Argentina (CONICET), Universidad Nacional de C\'ordoba, Laprida 854, X5000BGR, C\'ordoba, Argentina}
\affiliation{Observatorio Astronómico, Universidad Nacional de Córdoba, Laprida 854, X5000BGR, Córdoba, Argentina}

\author[0000-0002-2136-2591]{Laura Ceccarelli}
\affiliation{Instituto de Astronom\'ia Te\'orica y Experimental (IATE), Consejo Nacional de Investigaciones Cient\'ificas y T\'ecnicas de la\\ Rep\'ublica Argentina (CONICET), Universidad Nacional de C\'ordoba, Laprida 854, X5000BGR, C\'ordoba, Argentina}
\affiliation{Observatorio Astronómico, Universidad Nacional de Córdoba, Laprida 854, X5000BGR, Córdoba, Argentina}

\author[0000-0001-5262-3822]{Valeria Coenda}
\affiliation{Instituto de Astronom\'ia Te\'orica y Experimental (IATE), Consejo Nacional de Investigaciones Cient\'ificas y T\'ecnicas de la\\ Rep\'ublica Argentina (CONICET), Universidad Nacional de C\'ordoba, Laprida 854, X5000BGR, C\'ordoba, Argentina}
\affiliation{Observatorio Astronómico, Universidad Nacional de Córdoba, Laprida 854, X5000BGR, Córdoba, Argentina}

\author[0000-0002-2039-4372]{Facundo Rodriguez}
\affiliation{Instituto de Astronom\'ia Te\'orica y Experimental (IATE), Consejo Nacional de Investigaciones Cient\'ificas y T\'ecnicas de la\\ Rep\'ublica Argentina (CONICET), Universidad Nacional de C\'ordoba, Laprida 854, X5000BGR, C\'ordoba, Argentina}
\affiliation{Observatorio Astronómico, Universidad Nacional de Córdoba, Laprida 854, X5000BGR, Córdoba, Argentina}

\author[0000-0002-4480-6909]{Gian Luigi Granato}
\affiliation{INAF, Osservatorio Astronomico di Trieste, via Tiepolo 11, I-34131, Trieste, Italy}
\affiliation{Instituto de Astronom\'ia Te\'orica y Experimental (IATE), Consejo Nacional de Investigaciones Cient\'ificas y T\'ecnicas de la\\ Rep\'ublica Argentina (CONICET), Universidad Nacional de C\'ordoba, Laprida 854, X5000BGR, C\'ordoba, Argentina}
\affiliation{IFPU - Institute for Fundamental Physics of the Universe, Via Beirut 2, 34014 Trieste, Italy}

\author[0000-0002-4882-1735]{Andrea Lapi}
\affiliation{SISSA, Via Bonomea 265, I-34136 Trieste, Italy}
\affiliation{IFPU - Institute for Fundamental Physics of the Universe, Via Beirut 2, 34014 Trieste, Italy}
\affiliation{INFN-Sezione di Trieste, via Valerio 2, 34127 Trieste, Italy}
\affiliation{INAF/IRA, Istituto di Radioastronomia, Via Piero Gobetti 101, 40129 Bologna, Italy}

\author{Rien van de Weygaert}
\affiliation{Kapteyn Astronomical Institute, University of Groningen,PO Box 800, 9747 AD, Groningen, The Netherlands}


\begin{abstract}

The large-scale environment of the cosmic web is believed to impact galaxy evolution, but there is still no consensus regarding the mechanisms. We use a semi-analytic model (SAM) galaxy catalog to study the star formation and dust content of local galaxies in different cosmic environments  of the cosmic web, namely voids, filaments, walls, and nodes. We find a strong impact of the environment only for galaxies with $M_{\rm stars}\lesssim10^{10.8}\, M_\odot$: the less dense the environment, the larger the star formation rate and dust content at fixed stellar mass. This is attributed to the fact that galaxies in less dense environments typically feature younger stellar populations, a slower evolution of their stellar mass and a delayed star formation compared to galaxies in denser environments.  
As for galaxies with $M_{\rm stars}\gtrsim 10^{10.8}\, M_\odot$ differences among environments are milder due to the disc instability (DI) driven supermassive black hole (SMBH) growth implemented in the SAM, which makes SMBH growth, and thus galaxy quenching, environment insensitive. 
We qualitatively test our predictions against observations by identifying environments in the SDSS-DR16 using dust masses derived from the GAMA survey. The agreement is encouraging, particularly at ${\rm log} \, M_{\rm stars}/M_\odot\gtrsim 10.5-11$, where sSFRs and dust masses appear quite environment-insensitive. This result confirms the importance of in situ growth channels of SMBHs.

\end{abstract}



\section{Introduction}

\label{sec:intro}

Extensive redshift galaxy surveys, such as 2dFGRS \citep{2dFGRS}, SDSS \citep{York:2000}, GAMA \citep{gama} e\textsc{BOSS} \citep{eBOSS}, \textsc{DES} \citep{des}, have revealed that galaxies define a large scale cosmic web structure, ranging from vast low-density regions known as cosmic voids to higher-density regions such as walls, filaments, groups, and galaxy clusters.
This pattern is also present in the distribution of halos \citep{Bond1996, Cautun2014} in cosmological simulations, e.g. \textsc{EAGLE}  \citep{eagle_1, eagle_2},  \textsc{MultiDark} \citep{Klypin2016},  \textsc{IllustrisTNG} \citep{illustristng}. 
The cosmic web results from the gravity-driven evolution of primordial density perturbations. Hence, it embodies a large amount of information which may be exploited to test different cosmological models \citep[e.g.][]{Bos2012,Paillas21,Bonnaire22}. On the other hand, the properties of galaxies and their dependence on the cosmic web are crucial for understanding their formation and evolution. It is well established that galaxy local environment ($\lesssim 5 \, {\rm Mpc}$) strongly affects their star formation activity, colors and morphology \citep[e.g.][]{dressler80, Balogh2004,Kauffmann2004,Bamford:2009,Peng2010,Woo2013,Old2020}. These works show that galaxies tend to be less star-forming, redder and more elliptical as local density increases. 
On the other hand, galaxies and their local environments are embedded in different large-scale environments ($\gtrsim 10-100 \, {\rm Mpc}$) of the cosmic web (like cosmic voids, walls, filaments). However, the role of such environments on galaxy formation and evolution remains a matter of debate \citep[e.g.][]{Alfaro2022,Rodriguez-Medrano2023,Wang2023:halo}. In this work, we employ the term \textit{environment} to denote the large-scale environment, unless stated otherwise.\\

The densest regions of the cosmic web are nodes and filaments. The most massive nodes host galaxy clusters, which represent the largest entities in the Universe that exist in a state of quasi-virial equilibrium. They feature a deep gravitational potential well permeated by an intra-cluster medium (ICM) of hot ionized gas. In this dense and possibly hostile environment, multiple mechanisms influence galaxy evolution. Some of them, such as ram pressure stripping, strangulation and tidal stripping, trigger gas depletion, leading to the quenching of star formation (see e.g. \citealt{Abadi:1999,Vijayaraghavan:2015,Steinhauser:2016,Peng:2015,Gnedin:2003a}). 
Due to these processes and frequent interactions, clusters of galaxies host galaxies often characterized by red colours and elliptical morphologies \citep[e.g.][]{dressler80,Whitmore:1993,Dominguez:2001, Bamford:2009, Paulino-Afonso:2019}.

Nodes are connected by filaments, elongated structures which have undergone gravitational collapse along two principal axes. They represent a kind of \textit{bridges} along which matter flows to accrete into nodes. Filaments are the predominant visual features within the cosmic web and host galaxies which tend to exhibit a larger mass, a redder color, and earlier-type morphologies compared to their counterparts in less dense environments \citep{Chen2017,Laigle:2018,Kraljic2018,Kuutma17}.

In contrast to high-density regions, large voids represent the extremely low-density environment characterized by reduced galaxy mergers and interactions compared to the field or groups. Cosmic voids represent a unique and pristine environment where galaxies are unaffected by the transformation processes typical of over-dense galaxy systems, like clusters and groups. Thus, they allow to study galaxy evolution as a result of nature only, without nurture.

 Numerous studies have focused on galaxies in cosmic voids and found that they appear to have significantly different properties than field galaxies. The luminosity function of galaxies in voids (e.g. \citealt{Hoyle2005}) shows a fainter characteristic magnitude. However, the relative importance of faint galaxies is similar to that found in the field. Spectroscopic and photometric properties of void galaxies have also been studied in detail \citep{Rojas2005, Hoyle2005, Hoyle2012, Kreckel2012}. These results indicate that galaxies inside voids have higher star formation rates and bluer colours than galaxies in denser regions and are still forming stars at a rate relatively closer to the past one. Statistical studies using observational data report that void galaxies are smaller, bluer, later-type morphology and more star-forming than those in average density environment \citep{Grogin2000, Rojas2004, Hoyle2005, Patiri2006, Park2007, Wegner2008, Kreckel2011, Liu2015ApJ, Tavasoli2015, Moorman2016, Beygu2017, Ceccarelli2021, Jian2022}. Finally, we mention the relevant result recently presented by \cite{Dominguez23}, within the context of CAVITY project. The authors of the latter work performed a spectral analysis on a sample of nearby galaxies in voids, filaments and walls, and clusters. They find a clear correlation between the density of the large-scale environment and the star formation history (SFH), namely galaxies in less dense environments feature a slower SFH, that is, void galaxies assemble their stellar mass slower.\\

On the theoretical side, both hydrodynamic simulations and semi-analytic models (SAMs) have been used to shed light on the influence of the large-scale environment on galaxy evolution. \cite{Kreckel2011} identified higher star formation rates and younger stellar ages in void galaxies through their hydrodynamic simulation. \cite{Rieder13} studied the formation of (sub)structures in the halo distribution in voids, within the context of the Cosmogrid simulation. \cite{Habouzit2020} employed the Horizon-AGN simulation \citep{Dubois14}, coupled to the VIDE void finder \citep{Sutter15}, in order to investigate the black hole population within cosmic voids. They reported no significant variations in black hole growth in voids compared to more dense environments. \cite{Rosas-Guevara2022} studied central galaxies located within and near voids using the \textsc{EAGLE} simulation \citep{eagle_1,eagle_2} and the spherical voids catalog by \cite{Paillas17}. They analyzed in detail the star formation, metallicity, morphology and assembly history of galaxies as a function of their stellar mass and environment. \cite{Alfaro2020,Alfaro2021} studied the Halo Occupation Distribution (HOD) in cosmic voids and Future Virialized Structures (FVS). 
Making use of the semi-analytic catalog MDPL2-SAG \citep{knebe2018,Cora18} and the hydrodynamic simulation TNG300 \citep{illustristng}, they found a lower(higher) than average HOD and formation redshift in voids(FVS). Finally, \cite{Jaber2023} exploited the SAGE SAM \citep{Croton2016} and the Spine-Web algorithm \citep{AragonCalvo10} to study the large-scale dependence of the metallicity and stellar-to-halo mass ratio. Their results indicate the presence of a threshold mass (respectively $M_{\rm halo} \simeq 10^{12}\, M_\odot/h$ and $M_{\rm stars} \simeq 10^{10}\, M_\odot/h$) below which the stellar-to-halo ratio and metallicity are enhanced in dense environments.\\


Despite the substantial achievements of these studies, we still need a complete understanding of how galaxy properties correlate with the large-scale environment. The distinct underlying physical models within different simulations and the lack of a single, clear-cut definition for large-scale environments hinder a direct comparison of results. Furthermore, such complexities make it challenging to establish the relative impact of different physical mechanisms on galaxy evolution in distinct environments.\\


In this work, we study the dependence on the large-scale environment of the star formation and, for the first time, the dust content of local galaxies. We use a galaxy catalog obtained from a SAM, and different methods are adopted to identify cosmic environments, namely voids, walls, filaments, and nodes. The primary advantage of our approach lies in the computational efficiency of the SAM method, enabling us to simulate relatively large volumes. This capability is essential for obtaining a reliable statistical representation of large-scale cosmic environments. Furthermore, the simplicity of running the SAM, and the subsequent ability to test various physical models, makes easier the identification of the role of different physical processes in shaping the environmental dependence of galaxies properties.  
Also, adopting multiple methods to define environments enhances the reliability of our results and mitigates the inherent arbitrariness in defining large-scale structures.

In order to check our results, we compare them with observations. We use the Sloan Sky Digital Survey (SDSS-DR16) for environmental classification, stellar mass and star formation rates. At the same time, for a subset of galaxies, we utilize dust mass measurements from the GAMA and \textit{H}-ATLAS surveys. Our \textit{qualitative} comparison yields highly encouraging results, highlighting the significance of the in situ supermassive black hole (SMBH) growth mechanism, specifically secular accretion during disc instabilities (DIs) in our case. In a broader context, our work underscores the importance of investigating galaxy evolution in diverse environments as a possible way to assess the relative significance of in situ and ex situ processes.\\

The paper is organized as follows. The adopted SAM and the observational data are briefly described, respectively, in Sec. \ref{sec:sam} and \ref{sec:obs}. In Sec. \ref{sec:envclass} we describe the process of environment classification in both the simulated and observed samples. The star formation and dust content of simulated galaxies in different large-scale environments, as well as their evolution, are investigated in Sec. \ref{sec:galaxyprop} and \ref{sec:galaxyassembly}, while Sec. \ref{sec:SMBHgrowth} focuses on the impact of the DI-driven SMBH growth. In Sec. \ref{sec:compobs} we compare the SAM results with observations, and finally we draw our conclusions, after summarizing the work, in Sec. \ref{sec:discussionconclusion}.

\section{The semi-analytic model}
\label{sec:sam}
SAMs are valuable tools for studying galaxy evolution, owing to their exceptional capability to simulate extensive volumes while providing a comprehensive albeit simplified representation of numerous physical processes that shape the properties of galaxies \citep[e.g.][]{WhiteFrenk91,Cole2000}. SAMs provide simplified descriptions of baryonic processes shaping galaxy populations within Dark Matter (DM) halos merger trees. These processes include gas inflow, cooling, star formation, SMBH growth and feedback effects. The assumed sequence begins with gas collapse, forming rotation-supported gas disks with mild star formation. Spheroidal galaxies result from mergers and instabilities, potentially leading to starbursts. Also, environmental effects, e.g. ram pressure and tidal forces, are often considered. These baryonic processes are modeled through approximate and motivated relationships to evolve the galaxy population over time. 

In this work we adopt the last public release\footnote{The source code is available at \url{https://github.com/LGalaxiesPublicRelease/LGalaxies_PublicRepository/releases/tag/Henriques2020}} of the Munich galaxy formation model, \textsc{L-Galaxies} \citep{Henriques2020}, along with the updates introduced and discussed in \cite{Parente2023}, hereafter \citetalias{Parente2023}. In the latter work we implemented a detailed treatment of dust grains formation and evolution, as well as an updated treatment of disc instabilities (DIs). The dust model (Sec. 2.1 of \citetalias{Parente2023}) includes two sizes (large and small grains, with radii $0.05$ and $0.005 \, \mu{\rm m}$) and two chemical compositions (silicate and carbonaceous, MgFeSiO$_4$ and C) of grains. Briefly, grains are produced in AGB stars envelopes and type-II SNe ejecta, thus ejected into the surrounding gaseous medium. Here, the model takes into account different processes, which affect the mass (grains accretion, destruction in SNe shocks, thermal sputtering) and the size (shattering, coagulation) evolution of grains. These processes depend on the physical properties of the galaxy as provided by the SAM (e.g. molecular gas fraction, gas temperature, metallicity). As for the new treatment of DIs (Sec. 2.2 of \citetalias{Parente2023}), we now consider the instability of the combined gas+stars disc \citep[only the stellar disc was considered in][]{Henriques2020}. \rev{The disc becomes unstable when the centrifugal force cannot counteract its self-gravity. The instability is particularly relevant for building up the spheroidal component of low-to-intermediate mass (${\rm log} M_*/M_\odot \lesssim 10.5$) galaxies. In these episodes, the unstable stellar component is transferred into the bulge, while the unstable gas can both fuel a starburst and accrete onto the central SMBH. The fraction falling into the SMBH is $f_{\rm BH,\,DI}$, which is a decreasing function of the virial velocity of the halo, according to a popular phenomenological description used in SAMs \citep[][see Eq. 23 of \citetalias{Parente2023}]{Kauffmann00}. The chosen values of the free parameters in this relationship ensure that the model predictions align with the observed relative fraction of massive star-forming and quiescent galaxies (App. B of \citetalias{Parente2023}). For reference, $f_{\rm BH,\,DI}$ is of the order of $10^{-4}-10^{-5}$.}

The \rev{aforementioned} modifications allow our SAM to reproduce a broad range of galaxy properties concerning dust, in particular, the cosmic evolution of the galactic dust abundance and the local fraction of elliptical galaxies, still reproducing several crucial properties of the galaxy population (e.g. cosmic star formation, mass functions). Among the modifications introduced, the SMBH growth during DIs is particularly relevant since it actively operates to reduce star formation in the most massive objects due to the radio-mode AGN feedback, and it turns out to be very relevant for the results presented here (see extended discussion in Sec. \ref{sec:SMBHgrowth}). For this reason, along with the standard fiducial (FID) version of \citetalias{Parente2023}, in the following we will often discuss results obtained by switching off the SMBH growth channel during DIs ($f_{\rm BH,\,DI}=0$)\footnote{This means setting $f_{\rm BH}=0$ in Eq. 23 of \citetalias{Parente2023}.}.

Finally, regarding environmental processes, the adopted version of the \textsc{L-Galaxies} models tidal stripping, ram-pressure stripping, and tidal disruption, all processes that shape the properties of satellite galaxies. A detailed overview of these processes is available in the supplementary material of \citet[but see also \citealt{Ayromlou2021} for an improved treatment of environmental processes within the \textsc{L-Galaxies} framework]{Henriques2020}.\\

The SAM is run on top of the \textsc{Millennium} merger trees (\citealt{Springel05}, box size $500 {\rm Mpc}/h$, $2160^3$ particles), and a \textit{Planck} cosmology\footnote{The cosmology originally adopted in the \textsc{Millennium} simulation has been scaled according to \cite{Angulo2010} and \cite{Angulo2015}.} \citep{Planck14} is assumed throughout this work ($h=0.673$, $\Omega_{\rm m}= 0.315$, $\Omega_{\rm b}= 0.0487$, $\sigma_8=0.829$). A \cite{Chabrier2003} initial mass function (IMF) is adopted. We analyse galaxies with a stellar mass content of at least $\log(M_{\rm stars}/M_\odot) \geq 9$, which approximately corresponds to the resolution limit of the underlying DM simulation \citep[e.g.][]{Guo11}.

\section{Observational Data} \label{sec:style}
\label{sec:obs}

We exploit different data sets to test our model predictions concerning the star formation and dust content of galaxies in different environments. We use the Sloan Digital Sky Survey (SDSS) Data Release DR16 \citep{SDSS16} to identify cosmic environments. We use galaxies with $r-$band apparent magnitudes $r<17.77$ (completeness limit) and in the redshift range $0.02<z<0.1$. We take stellar masses and star formation rates from the MPA-JHU catalog \citep{Kauffmann2003}.

As for dust masses, we exploit the results presented in \cite{Beeston18}, who studied the local ($z<0.1$) dust mass function for galaxies in the GAMA/\textit{H}-ATLAS surveys. Specifically, the physical properties of these galaxies (including dust mass) were obtained by \cite{Driver18} using a spectral energy distribution (SED) fitting procedure.

\section{Environmental classification}
\label{sec:envclass}

We employ various methods to identify cosmic environments within the simulated and observed large-scale structures. We categorize environments in two fundamental ways: one involves employing a single and homogeneous method, \NEXUS{}, to segment the cosmic web into voids, walls, filaments and nodes. \rev{The other approach, which will be often dubbed as R19+T23+FOF, utilizes multiple independent methods for identifying specific environments (voids and walls, filaments, and groups).} The former is exclusively applied to the simulation, while the latter is employed in both the observed and simulated galaxy catalogs. \rev{It is important to stress that no tuning of their parameters has been made to match environments extracted from \NEXUS{} to those extracted with the R19+T23+FOF approach.} \rev{As can be appreciated in Fig. \ref{fig:maps}, there are various differences between these two approaches, the most important one being the number of galaxies detected as belonging to a given environment. While \NEXUS{} associates \textit{every} galaxy in our simulation to a given environment, this is not true for the R19+T23+FOF method. However, although there is a discrete visual match between the environments detected by the two approaches, we stress the objective of our study is not to compare the performances of different identification algorithms.} Instead, our focus is on examining the consistency and reliability of our results concerning the influence of the environment on galaxy properties, regardless of the method used to identify the different environments. 

In the following sections, we introduce each specific environment we focus on and a comprehensive overview of each identification method adopted.

\label{sec:envclass:sam}
\subsection{Cosmic web segmentation: \NEXUS{}}
\label{sec:envclass:sam:NEXUS}

For the segmentation of the simulated cosmic web into its distinct structural environments we use the Multiscale Morphology Filter (MMF)/NEXUS pipeline, specifically its \NEXUS{} version \citep{Cautun2013,Cautun2014}.  It is the highest dynamic range version of the library of routines for pattern classification based on the MMF/NEXUS formalism \citep[][see also \citealt{Libeskind18} for a short review]{Aragon2007,AragonCalvo10,Cautun2013,Cautun2014,Aragon2014}. This formalism represents a scale adaptive framework that classifies the matter distribution on the basis of local spatial variations in either the density field, velocity field or gravity field. Subsequently, a set of morphological filters is used to classify the spatial matter distribution into three basic components, the clusters, filaments and walls that constitute the cosmic web. The remaining volume elements are classified as part of voids. The end product of the pipeline is a a map in which for each location in the analyzed volume the morphological identity is specified.


In practice, \NEXUS{} takes as input a regularly sampled density field, which we obtain by projecting the halo distribution to a regular grid using the cloud-in-cell interpolation scheme. In a first step, the input field is Gaussian smoothed using a Log-density filter that is applied over a set of scales from $1$ to $11h^{-1}\mathrm{Mpc}$. It results in a map of $1024^3$ cells, corresponding to a cellsize of $0.472h^{-1}\mathrm{Mpc}$. Then, for each smoothing scale, the resulting density is used to calculate the Hessian matrix and its corresponding eigenvalues, whose values and signs determine the environment response at each location, i.e. grid cell. In the last step, \NEXUS{} combines the environmental signature of all the scales to obtain a scale-independent value. In this work, roughly $79.3\%$ of the total volume of the box corresponds to cells located in voids, $14.5\%$ is classified as a belonging to a walls, $6.0\%$ as located in filaments and a mere $0.2\%$ correspond to nodes. Assigning to each halo the environment of it hosting cell, we find that, out of the total DM mass within the box, $9.5\%$ resides in voids, $16.9\%$ in walls, $34.5\%$ in filaments and $39.1\%$ in the cluster nodes of the cosmic web.

\subsection{Spherical voids identification}

\label{sec:envclass:voids}
We use the spherical void finder algorithm presented in \citet{Ruiz2015,Ruiz19} as an alternative method to construct our voids catalog. This method aims at identifying the largest and non-overlapping spherical regions that satisfy that the number density of tracers inside them is less than 10 percent of the mean:
\begin{equation}
\Delta(r_{\rm void}) \le -0.9,     
\end{equation}
where $\Delta(r)$ is the integrated density contrast at scale $r$ and $r_{\rm void}$ is the void radius.

We use the DM halos as structure tracers when applying this algorithm to our SAM galaxies. We adopt $M_{\rm halo} \ge 10^{12} \hmsun$ as halo mass cut, obtaining 2729 voids. After constructing the voids catalog, we define void galaxies as those residing within $r/r_{\rm void}\leq1$ and void wall galaxies as those located immediately outside the voids with $1< r/r_{\rm void} \leq 1.2$. \rev{These void galaxies are predominantly found in \NEXUS{} voids ($\sim 64\%$), while galaxies in the voids shells are more homogeneously distributed among \NEXUS{} voids, walls and filaments (respectively, $\sim 31\,\%,\,34\,\%,\,32\,\%$).}\\

The same algorithm is adopted to identify voids on top of the observational catalog. Details about the algorithm application to observational samples are provided in \citet{Ruiz19}. Specifically, in this work the void identification is the same as in \citet[][see their Sec. 2.1]{Rodriguez-Medrano2023}. Voids are identified in a volume complete sample with limiting redshift $z=0.1$ and maximum absolute magnitude in the r band $M_{\rm r}-5 {\rm log}h=-20$ extracted from the main SDSS region. In our analysis, we have discarded voids close to the edge of the catalog, namely, void centers at distances smaller than $1.8  r_{\rm void}$. Starting from this voids catalog we then define void ($r/r_{\rm void}\leq1$) and void wall galaxies ($1<r/r_{\rm void}\leq1.2$).

\subsection{Filaments and groups identification}

\label{sec:envclass:filsgroups}

As for the more dense environments, we also identify galaxy groups and filaments with an alternative approach. 
Galaxy groups are identified with \rev{Friends-Of-Friends (FOF)} haloes in our SAM catalog. As for observed galaxies, we utilize the groups catalog presented by \citet{Rodriguez22}, which is derived from the SDSS-DR16 using the algorithm introduced in \citet{RodriguezMerch20}.
This method initially applies the FOF algorithm to identify galaxy systems that are gravitationally bound and have at least one bright galaxy with an r-band absolute magnitude brighter than -19.5. Subsequently, a halo-based algorithm \citep{Yang05,Yang07} is applied. A three-dimensional density contrast is calculated in redshift space using a characteristic luminosity calculated with the potential FOF galaxy members. The estimation procedure considers the incompleteness caused by the observational catalog's limiting magnitude. Next, by abundance matching on luminosity, the mass of each group is assigned. Assuming that galaxies populate the DM halos following a \cite*{NFW} profile and using the assigned mass, the three-dimensional density contrast is calculated to associate galaxies to the groups. With this last assignment of members, the procedure recalculates the characteristic luminosity and iterates until it converges. This algorithm showed excellent results in purity and completeness \citep{RodriguezMerch20}.\\

Once galaxy groups are identified, they are used to build a catalog of filaments in the large scale structure. They are cylinders linking groups of galaxies. Our approach is similar to the filament identification carried out by \citet{Taverna:2023}, which, in turn, is based on the original algorithm by \citet{Martinez16}. Starting from the group sample, we search for all pairs of groups more massive than a selected mass cut-off $M_\mathrm{min}$ that are separated (in redshift space in the case of the observations) by a comoving distance smaller than a given threshold $\Delta_\mathrm{max}$. For each pair of groups thus selected, we compute the numerical overdensity of galaxies in a cylinder of radius $R$ that extends between the two groups. The overdensity in this volume is obtained by the ratio between the number of galaxies that lie in the cylinder, and the number of points within the same cylinder from an unclustered homogeneous distribution of points. This homogeneous distribution of points mimics the selection function of galaxies while being 100 times denser in order to reduce shot noise in the overdensity computation. We require the numerical overdensity of galaxies in 
the cylinder to be greater than unity in order to consider a pair of groups are linked by a filament. Galaxies in these cylindric regions are considered to be galaxies in filaments.

In this work we use $M_\mathrm{min}=10^{13.5}\,M_{\odot}$, $\Delta_\mathrm{max}=20\ $Mpc, which is the correlation length of groups more massive than $M_\mathrm{min}$ according to \citet{Zandivarez:2003}, and $R=2.2\ $Mpc (see \citealt{Taverna:2023}). \rev{Galaxies identified in cylindric filaments belong mainly to \NEXUS{} filaments ($\sim55\%$), walls ($\sim25\%$), and nodes ($\sim 19\%$), while galaxies in massive groups almost uniquely belong to \NEXUS{} nodes ($\sim 95\%$).} For the SDSS-DR16 groups and galaxies, our unclustered homogeneous distribution of points mimics the selection function of SDSS-DR16 galaxies regarding angular coverage and redshift distribution.

\section{Properties of galaxies across cosmic environments}


\begin{figure*}[htb]
    \centering
    \includegraphics[width=0.9\columnwidth]{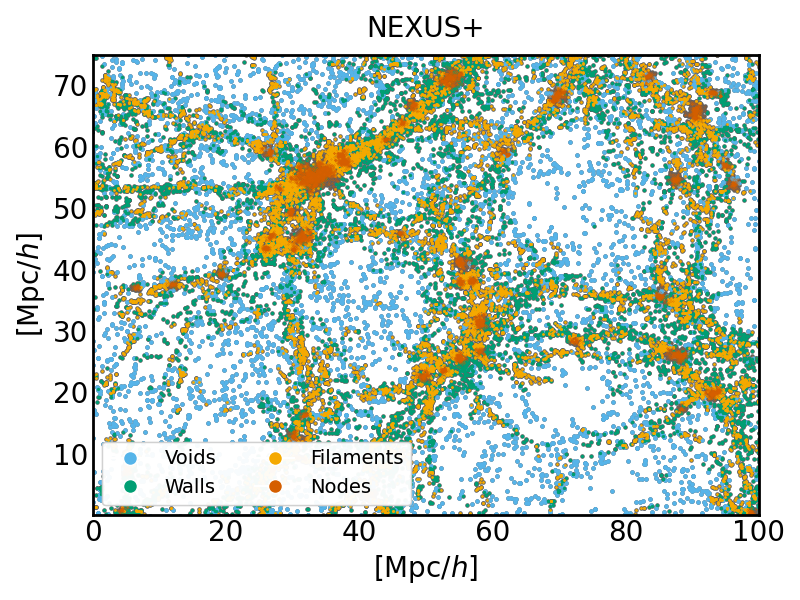}\quad
    \includegraphics[width=0.9\columnwidth]{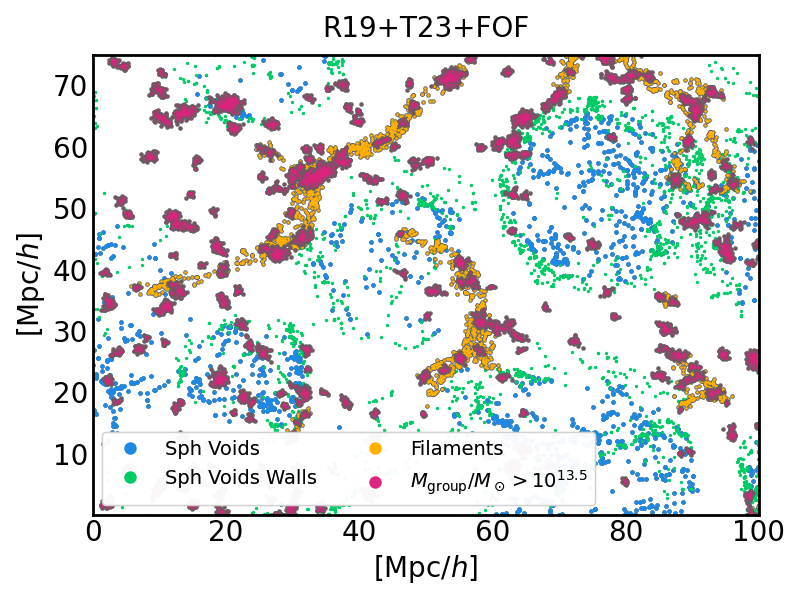}
    \caption{A $25 \, {\rm Mpc}/h$ slice of our simulation box showing the outcome of the environment identification processes. Each point represents a DM halo, and is color-coded according to the assigned environment. The left panel shows the results obtained by \NEXUS{}, which categorizes \textit{each} DM halo of the simulation as belonging to a void, wall, filament, or node. On the right panel we show the combination of the specific identifiers for distinct environments (R19+T23+FOF), i.e. spherical voids and their walls, cylindrical filaments, and massive groups ($M_{\rm group}>10^{13.5}\, M_\odot$).  It is important to note that \rev{while \NEXUS{} assigns every DM halo to an environment, this is not necessarily the case for the R19+T23+FOF approach, where a galaxy might not be classified as belonging to any specific environment.}}
    \label{fig:maps}
\end{figure*}

\begin{table}
\begin{center}
     \begin{tabular}{c c c} 
 \hline
 \hline
 {} & \multicolumn{2}{c}{Number of objects $[10^5]$} \\
 {} & {\NEXUS{}} & {R19+T23+FOF}\\
\hline\hline
Voids & {$33.6 $} & {$9.3 $} \\
Walls & {$38.4 $} & {$19.0 $}\\
Filaments &  {$49.9 $} & {$8.6 $} \\
{Nodes / Massive groups} &  {$30.1 $} & {$11.0 $}\\
\midrule[0.2\arrayrulewidth]

\hline\hline
\end{tabular}
\caption{\rev{Number of galaxies in our simulated catalog categorized based on environmental classification via the two approaches employed in this work, namely \NEXUS{} and R19+T23+FOF (Sec. \ref{sec:envclass}).}}
\label{tab:N_objs_SAM}

\end{center}

\end{table}

This section examines various physical properties of our simulated galaxies in different environments. Namely, we look at the star formation and dust content. The environmental dependence of dusty properties, such as the Dust-to-Gas (DTG) ratio, the size and chemical composition of dust grains, is discussed in App. \ref{sec:dustprop}. 

In the following analysis, we show the results for the \NEXUS{} environments (voids, walls, filaments and nodes), as well as for the other environments definitions adopted in this work, namely spherical voids and their associated walls, cylindrical filaments and massive groups ($M_{\rm group}>10^{13.5}\, M_\odot$). A visual representation of the environments identification in our simulated box can be appreciated in Fig. \ref{fig:maps} for both the aforementioned methods\rev{, and we report the number of objects identified in each environment in Tab. \ref{tab:N_objs_SAM}}.

\subsection{Star Formation and Dust}
\label{sec:galaxyprop}
The environment selection introduces a mass bias, namely, more massive galaxies are associated with more clustered environments (App. \ref{app:SMFsat}). To avoid this issue, we could study the properties of galaxies with identical stellar mass distribution in different environments. The procedure of selecting samples with the same mass distribution as in voids certainly removes the mass bias, however, since void galaxies are typically less massive than the whole population, this analysis may not be representative of the full range of $M_{\rm stars}$. For this reason, in this section we analyze galaxy properties in different environments \textit{at fixed stellar mass}.

\begin{figure*}[htb]
    \centering
    \includegraphics[width=0.8\columnwidth]{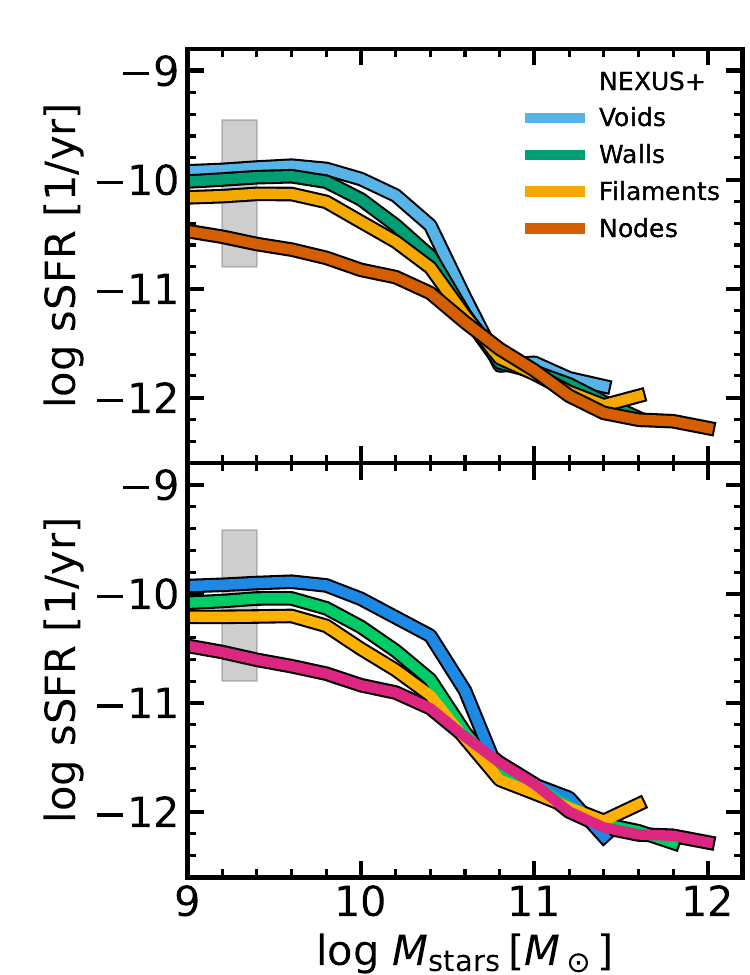}\qquad\qquad
    \includegraphics[width=0.8\columnwidth]{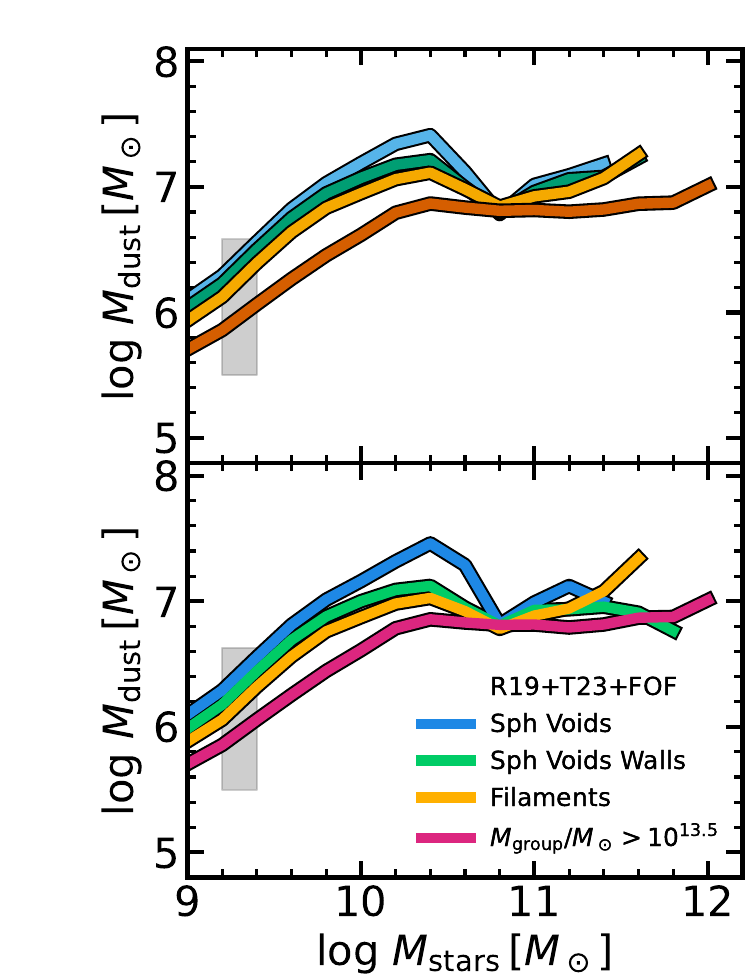}
    \caption{Specific SFR (\textit{left panel}) and dust mass (\textit{right panel}) as a function of stellar mass for simulated galaxies in different environments. We report the results for \NEXUS{} environments (voids, walls, filaments and nodes, \textit{top panels)}, as well as for the other environments definitions adopted in this work, namely spherical voids and their walls, cylindrical filaments and massive groups (R19+T23+FOF, \textit{bottom panels}). Solid lines represent median trends, with the typical $16-84$th percentile dispersion shown as a gray shaded area.}
    \label{fig:scaling_SF_dust}
\end{figure*}

We show in Fig. \ref{fig:scaling_SF_dust} the specific Star Formation Rate (sSFR) and dust mass as a function of stellar mass. In the range $9 \leq {\rm log}\,  M_{\rm stars}/M_\odot \lesssim 10.8$ we observe a clear trend with the environment, regardless of the environment definitions adopted. Namely, at fixed stellar mass, the \textit{less dense} is the environment, the more star-forming and dust-rich are the galaxies. Above this stellar mass value, we observe no clear trend with the environment for the sSFR. As for the dust mass, differences among environments are strongly reduced. This result suggests that the environmental dependence of dust and star formation properties is suppressed in galaxies above the stellar mass threshold. We will discuss this point at length in the following. 

We note that our finding of higher star formation in voids is in keeping with some observational results which pointed out that void galaxies are typically more SF and gas-rich than non-void ones (e.g. \citealt{Rojas2005}; \citealt{Benda-Beckmann2008}; \citealt{Kreckel2014}; \citealt{Beygu2016}; \citealt{Moorman2016}; \citealt{Florez2021}). The same has also been found in numerical simulations (e.g. \citealt{Cen2011}; but see also \citealt{Rosas-Guevara2022}, who found different environmental trends in different stellar mass regimes). Our findings are also consistent with \citet{Martinez16} particularly in relation to the sSFR of filament galaxies. These rates are intermediate falling between those observed in groups and those in less dense environments. They additionally find indications that variations among environments are less pronounced when considering high mass galaxies.\\


\subsection{Galaxy evolution across environments}
\label{sec:galaxyassembly}

\begin{figure}

\centering
\includegraphics[width=0.8\columnwidth]{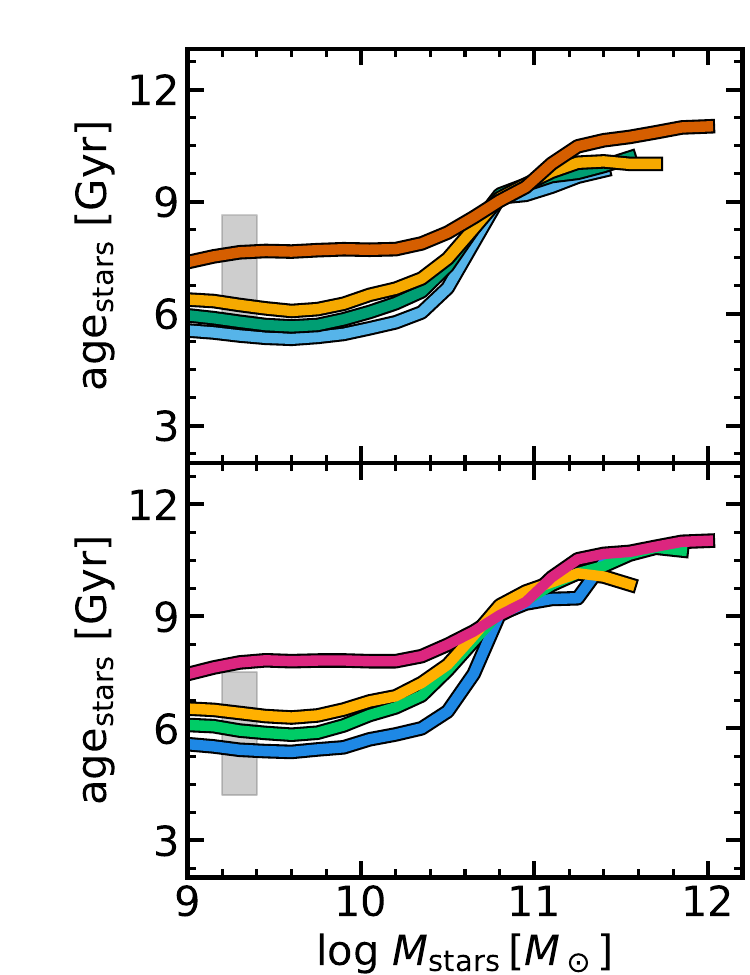}
\caption{Mass-weighted stellar populations age as a function of stellar mass for simulated galaxies in different environments, as detailed in Fig. \ref{fig:scaling_SF_dust}.}
\label{fig:age} 
\end{figure}

The scenario depicted above may be understood if galaxies undergo a \textit{slower} evolutionary process in \textit{less dense} environments. 
This idea is corroborated by the mass-weighted age of stellar populations, displayed in Fig. \ref{fig:age}, which suggests the existence of a relationship between age and environment for $M_{\rm stars}\lesssim 10^{10.8}\, M_\odot$. Voids galaxies are the youngest and nodes (or massive groups) galaxies exhibit the oldest ages. Walls and filaments galaxies fall in between, with filament galaxies systematically displaying older ages compared to wall galaxies.
We devote this section to further investigation of this point.
In particular, we aim to understand the differences in the evolution of galaxies with equal stellar mass at $z=0$, but residing in different environments.

\begin{figure*}[htb]
    \centering
    
    \includegraphics[width=2\columnwidth]{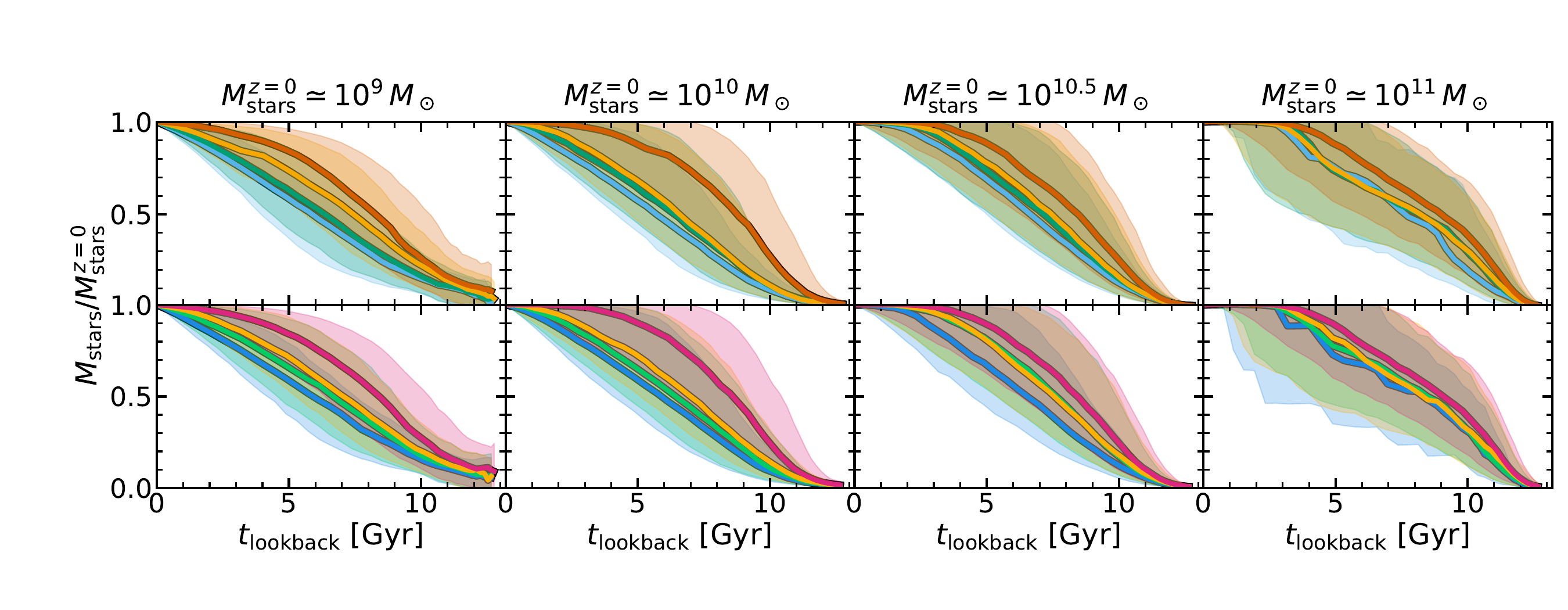}\\

    \includegraphics[width=2\columnwidth]{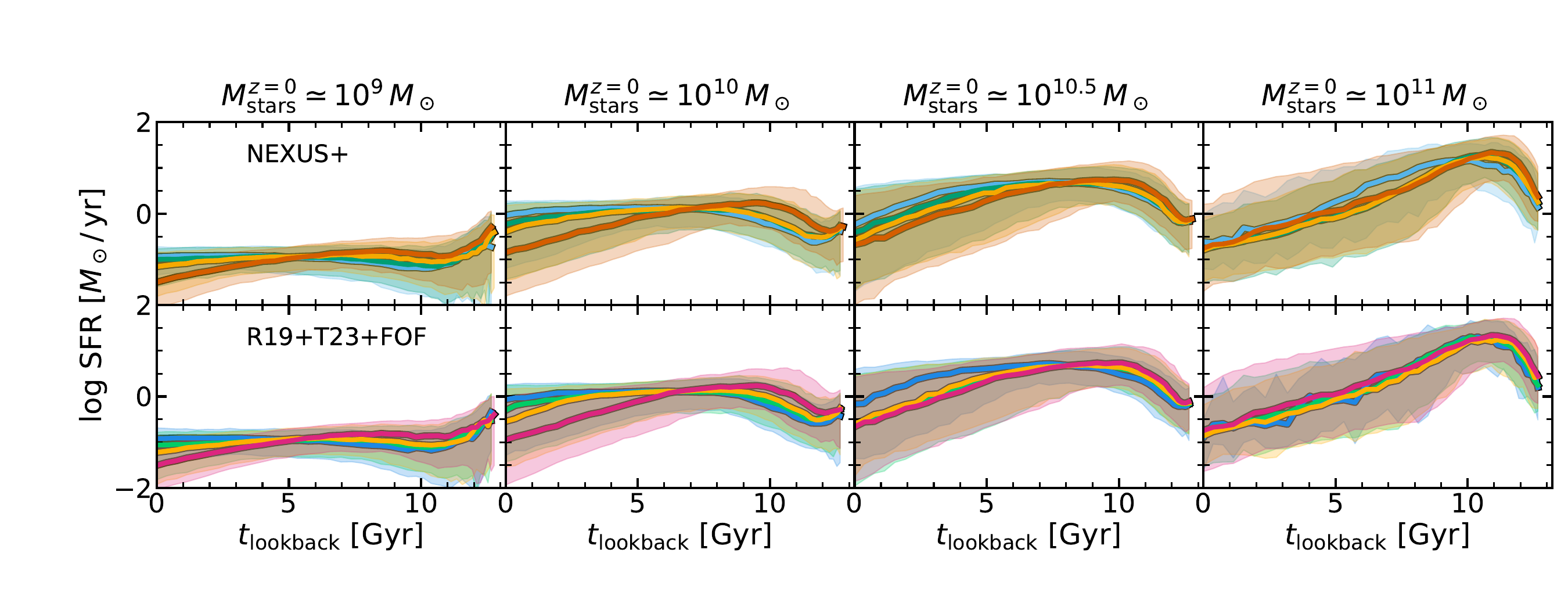}\\

    \includegraphics[width=2\columnwidth]{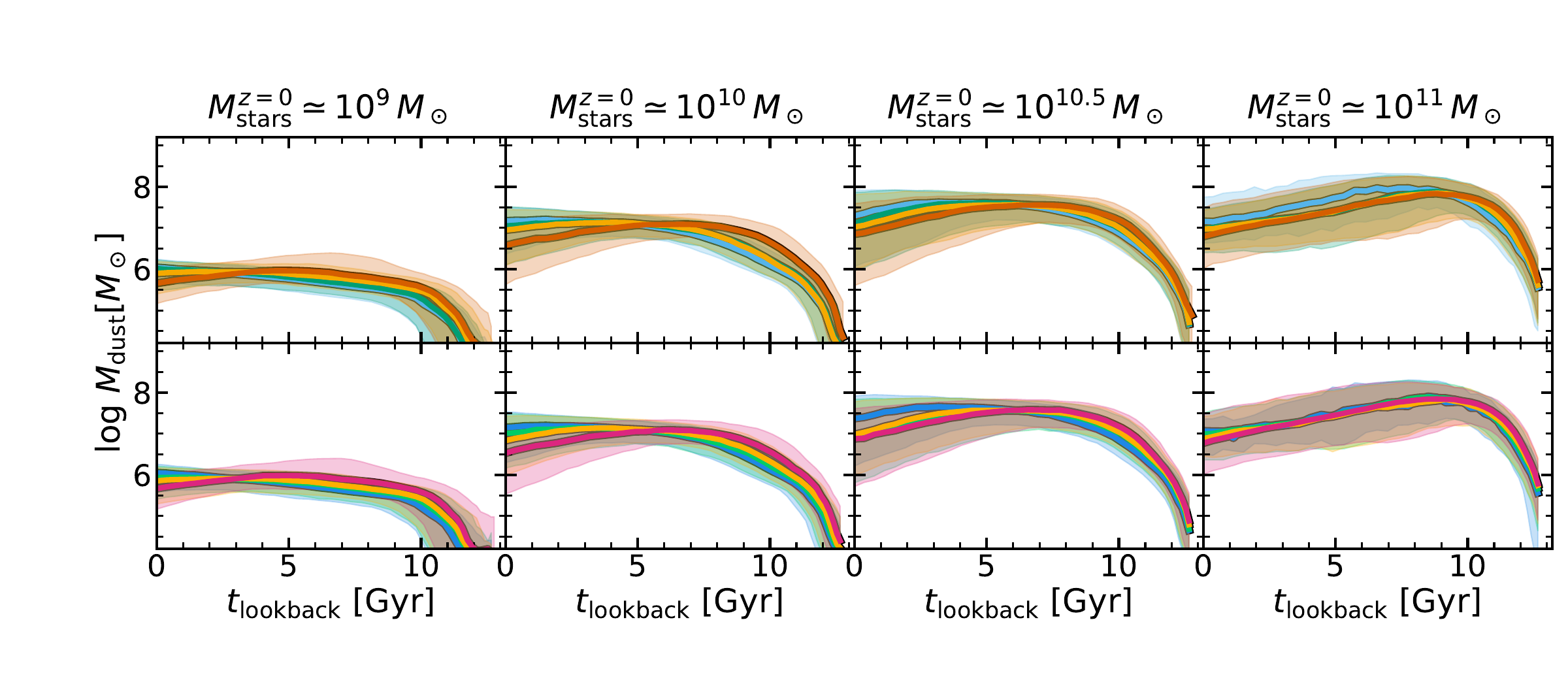}\\
    
    \caption{Stellar mass (\textit{top panels}), SFR (\textit{middle panels}), and dust mass (\textit{bottom panels}) evolution for galaxies with different values of the present-day stellar mass, namely (from left to right) $M_{\rm stars}^{z=0}/M_\odot \simeq 10^9, \, 10^{10},\, 10^{10.5},\, {\rm and}\, 10^{11}$.  
    For each quantity, we report the results for galaxies residing at $z=0$ in different \NEXUS{} environments (voids, walls, filaments and nodes), as well as for the other environments definitions adopted in this work, namely spherical voids and their walls, cylindrical filaments and massive groups (R19+T23+FOF). Colors are as in previous figures. For each environment, we randomly selected $\sim 10^3$ objects with $M_{\rm stars}^{z=0}$ in a narrow range ($0.1 \, {\rm dex}$) around the reported value, and traced their evolutionary paths along the merger tree back in time. Solid lines represent median trends, with the $16-84$th percentile dispersion shown as a shaded area.}
    \label{fig:evo:mstars}
\end{figure*}

We thus inspect the stellar mass\footnote{Normalized to the $z=0$ value.}, SFR, and dust mass evolution for galaxies with $M_{\rm stars}^{z=0}/M_\odot \simeq 10^9, \, 10^{10},\, 10^{10.5},\, {\rm and}\, 10^{11}$ in Fig. \ref{fig:evo:mstars}. The figures indicate that the \textit{denser} is the environment, the \textit{faster} is the evolution for $M_{\rm stars}^{z=0}\leq 10^{10.5} \, M_\odot$, in the sense that the SFR peaks at earlier epochs, and consequently the evolution of the stellar mass is more rapid. This holds true irrespective of the chosen environments definition.
As for the dust mass evolution, it reaches a maximum more recently than the SFR, suggesting that the time of the maximum dust content of a galaxy does not correspond to the time of the maximum of its star formation activity. We argue that this is due to the fact that the dust budget is determined not only by the production of dust by stars but also by its evolution, and by grains accretion in particular\footnote{The efficiency of the accretion process is delayed with respect to the dust production by stars also on a cosmic scale, as can be appreciated in Fig. 11 of \citetalias{Parente2023}.}. We also note that the peak of the dust abundance occurs later in lower density environments \textit{and} in lower $M_{\rm stars}^{z=0}$ objects. This maximum has yet to be reached in voids and wall galaxies of the less massive $M_{\rm stars}^{z=0}$ bin. This is consistent with the slower evolution of galaxies in lower density environments.

The slower stellar assembly in less dense environments has already been pointed out by other works based on hydrodynamical simulations \citep{Artale18,Alfaro2020,Habouzit2020,Rosas-Guevara2022}, and recently confirmed through observations by \cite{Dominguez23}.\\

As for the most massive stellar bin under examination, galaxies with $M^{z=0}_{\rm stars} \simeq 10^{11} M_\odot$ show no notable distinctions across diverse environments. The lack of differences in the evolution of these high-mass galaxies aligns with the findings discussed in Sec. \ref{sec:galaxyprop}, which suggest the existence of a certain stellar mass threshold ($\simeq 10^{10.8}\, M_\odot$) above which the influence of the large-scale environments becomes less relevant. 

\begin{figure}

\centering
\includegraphics[width=0.8\columnwidth]{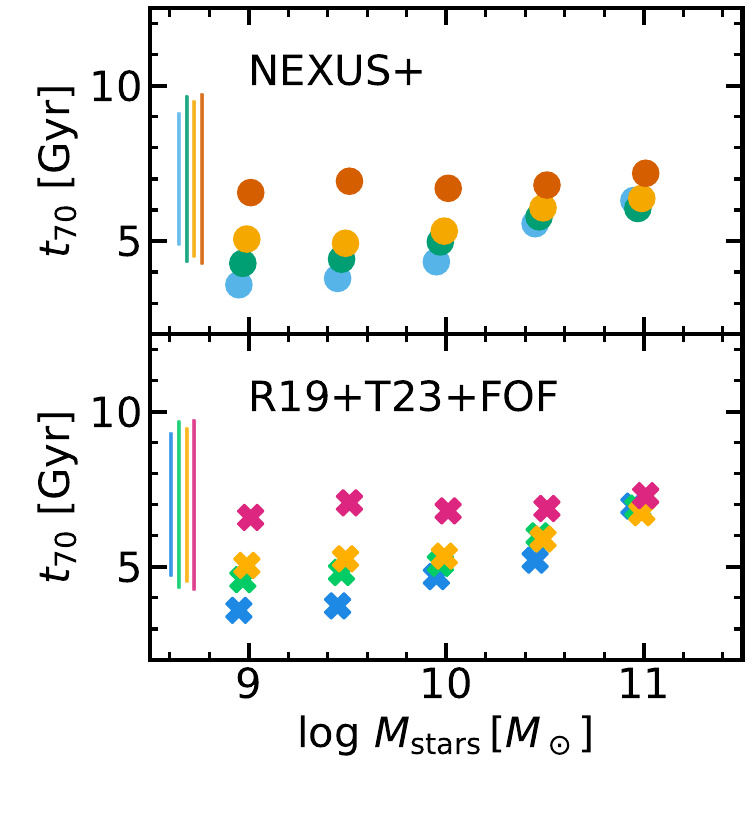}
\caption{\rev{Assembly time of the stellar mass for galaxies in different environments and different $z=0$ stellar mass bins. The time is expressed as $t_{70}$, the look-back time when the galaxy reached the $70\%$ of the present stellar mass. Crosses and circles refer to median values, while the typical $16-84$th percentile dispersion is reported as lines on the left side of each panel. Colors and environments are as in Fig. \ref{fig:scaling_SF_dust}.}}
\label{fig:t70} 
\end{figure}

\rev{The same conclusions may be drawn from an analysis of $t_{70}$, the look-back time at which the stellar mass of a given galaxy is $70\%$ of its present day value. In Fig \ref{fig:t70} $t_{70}$ is shown as a function of stellar mass and for different environments. First, we note that for galaxies with ${\rm log}\,M_{\rm stars}/M_\odot\gtrsim 9$,   $t_{70}$ tends to increase with stellar mass for voids, walls, filaments and, more weakly, nodes and massive groups. The last is due to the significant impact of old satellite galaxies, mainly dominating the population of these dense environments at ${\rm log}\,M_{\rm stars}/M_\odot\lesssim10.5$. Secondly, we observe that for ${\rm log}\,M_{\rm stars}/M_\odot\lesssim10.5-11$, at a fixed stellar mass, the time of assembly for the stellar mass is delayed in less dense environments. This delay is reflected in $t_{70}$, which is typically smaller for galaxies in less dense environments. Above the mentioned mass, there is no clear dependence on the environments, as already pointed out earlier.\\}


\rev{Notably, the observational study by \cite{Dominguez23} also revealed minimal changes in the assembly time of high-mass galaxies (log $M_{\rm stars}/M_\odot \simeq 10.5-11$) across various environments, contrasting with galaxies of smaller masses ($9<M_{\rm stars}/M_\odot \lesssim 10.5-11$, refer to their Fig. 4a and 4d). They hypothesized that the evolution of these high-mass galaxies might be more influenced by local interactions or their massive DM halos than by large-scale environments. In our model, we anticipate that, regardless of galaxy interactions, the in-situ growth of SMBHs in galaxies is crucial for understanding this behavior. Likewise, \cite{Dominguez23} observed similar assembly times for galaxies in their lowest stellar mass bin (log $M_{\rm stars}/M_\odot < 9$), suggesting a strong impact of the small-scale environment on objects of these masses, which are predominantly satellites. Since our analysis is limited to galaxies with log $M_{\rm stars}/M_\odot \geq 9$, we cannot determine if this behavior is also present in our model.}

\subsection{The impact of in situ SMBH growth on isolated galaxies evolution}

\label{sec:SMBHgrowth}


\begin{figure*}[htb]
    \centering
    \includegraphics[width=0.8\columnwidth]{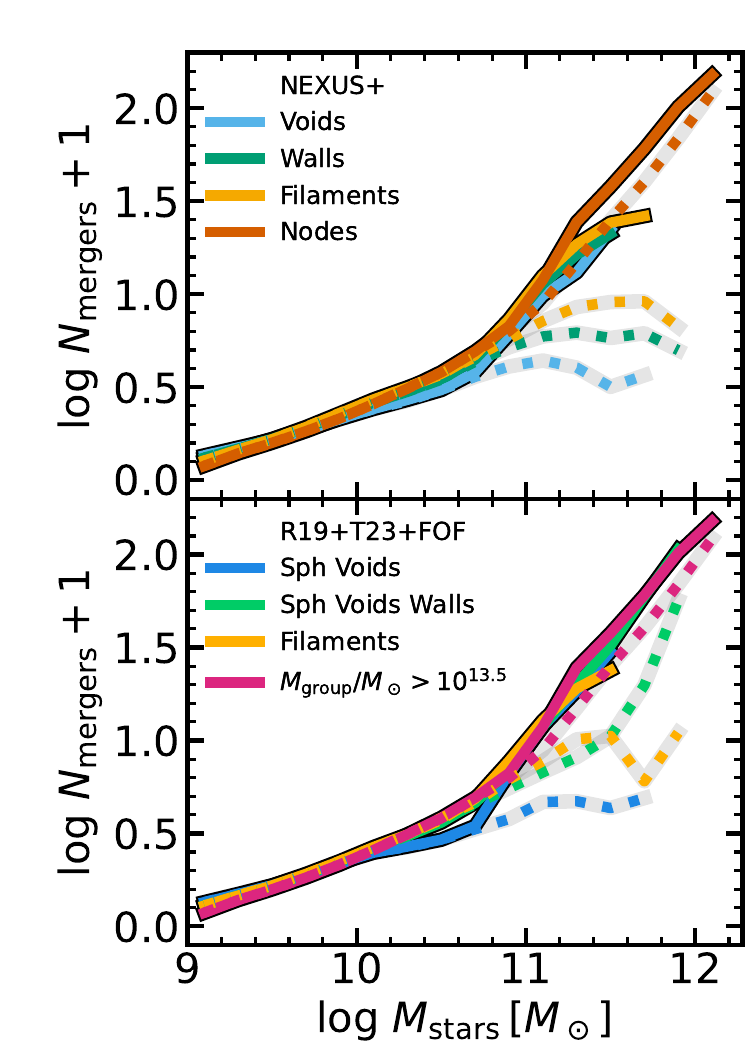}\quad
    \includegraphics[width=0.8\columnwidth]{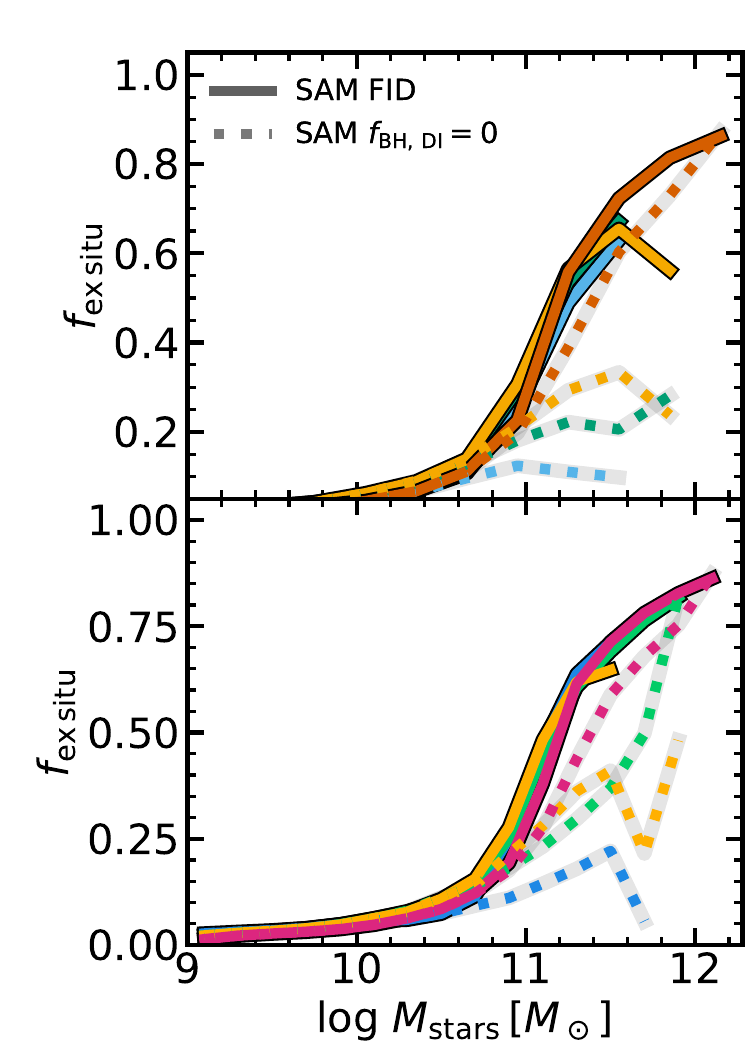}
    \caption{Mean number of mergers (\textit{left panel}) and ex situ fraction of the total stellar mass (\textit{right panel}) as a function of stellar mass for simulated galaxies. We report the results for galaxies residing at $z=0$ in different \NEXUS{} environments (voids, walls, filaments and nodes, \textit{top panels}), as well as for the other environments definitions adopted in this work, namely spherical voids and their walls, cylindrical filaments and massive groups (R19+T23+FOF, \textit{bottom panels}). Results from both the fiducial (\textit{solid}) and $f_{\rm BH, \, DI}=0$ (\textit{dotted}) SAM are shown.}
    \label{fig:assembly}
\end{figure*}

The latter result(s) is quite interesting. It suggests that the environmental dependence of galaxy evolution is important up to some characteristic mass (log $M_{\rm stars}/M_\odot \simeq 10.8$ in our model). Above this \textit{threshold}, galaxies in different large-scale environments and the same stellar mass feature very similar properties.

This outcome is due to the treatment of SMBH growth and feedback, the main channel of massive galaxies quenching in our SAM. More in detail, in our model, SMBHs can grow during mergers, by hot gas accretion, and during disc instabilities (DIs). \rev{In massive galaxies, the contribution of the latter process\footnote{It has been introduced in \citetalias{Parente2023}, while the other two channels are left as in the public \textsc{L-Galaxies} release \citep{Henriques2020}.} is typically less relevant than mergers in terms of mass. However, this channel is crucial in isolated systems, where (in absence of mergers) it is the only available channel for starting the SMBH growth, which will eventually quench the galaxy. Thus it has a profound impact on the results discussed in this work.} 

We demonstrate this by comparing the results obtained by our fiducial (FID) SAM and a version in which the SMBH growth in DIs is switched off ($f_{\rm BH,\,DI}=0$). In particular, Fig. \ref{fig:assembly} displays the number of mergers experienced up to $z=0$ and the fraction of stellar mass assembled ex situ (i.e. in mergers) for galaxies in distinct environments for both models. In the $f_{\rm BH,\,DI}=0$ model there are evident differences across environments for ${\rm log} \, M_{\rm stars}/M_\odot \gtrsim 10.8$, namely the denser the environment, the larger the number of mergers experienced at $z=0$, and consequently the fraction of stellar mass assembled ex situ. However, these differences are not present in the FID model. At first glance, this is quite surprising since we would expect the number of mergers to be independent of the baryonic physics implemented within the model, as it is determined by the adopted DM halos merger tree\footnote{Actually, the number of mergers experienced by a galaxy up to a certain redshift also depends on the time a galaxy need to merge once its associated DM subhalo merged. In our SAM, such a \textit{merging time} is estimated with the dynamical friction timescale \citep{Binney87}, as detailed in Sec. S1.16.1 of the supplementary material of \cite{Henriques2020}.}.

The reason why our novel\footnote{We clarify that the SMBH growth during DIs is not a novelty in the panorama of SAMs, since it has been adopted by the community for a long time \citep[e.g.][]{Croton2016, Lacey2016, PLagos2018}, also in the context of the same \textsc{L-Galaxies} model \citep{Irodotou2019,Izquierdo-Villalba2020}. The wording \textit{novel} adopted here has to be intended as \textit{novel with respect to the public release} of the \textsc{L-Galaxies} SAM \citep{Henriques2020}.} in situ SMBH growth channel produces such differences may be understood in the following picture. The DI-driven SMBH growth enhances the quenching of the most massive objects (\citetalias{Parente2023}), thus only galaxies that can accumulate a significant mass fraction through mergers are capable of reaching large stellar masses (${\rm log} \, M_{\rm stars}/M_\odot \gtrsim 10.8$), since the in situ star formation is suppressed by the instability-driven SMBH growth. This effect is particularly pronounced in galaxies located in less dense environments, which generally undergo fewer merger events. This is the reason why also the fraction of stellar mass formed through mergers, i.e. acquired ex situ, is influenced by our DI model (Fig. \ref{fig:assembly}).\\

In other words, according to our fiducial SAM, \textit{only galaxies which acquire a substantial stellar mass from mergers can reach large} $M_{\rm stars} \, (\gtrsim 10^{10.8}\, M_\odot)$.\\

Concluding this section, it is worth remarking that our DI model significantly modifies the in situ evolution of galaxies\footnote{Nonetheless, we note that our model still reproduces many crucial properties of the galaxy population, including the Stellar Mass Function, which is only mildly affected by the DI model (Sec. 3.1 \citetalias{Parente2023}).}, particularly impacting the environmental dependence of certain properties. This is not surprising, considering mergers' prominent role in shaping galaxy evolution within SAM frameworks. Consequently, we speculate that a thorough (also observational) study of galaxy properties in different cosmic environments could serve as an effective test bench for assessing the relative significance of the in situ and ex situ processes implemented within SAMs. We perform a step in this direction in Sec. \ref{sec:compobs}, where we compare the specific SFR and dust content for the FID and $f_{\rm BH, \, DI}=0$ models across different large-scale environments with observations.

Finally, it is noteworthy that \cite{Jaber2023}, using the SAGE SAM, observed a significant increase in metallicities for galaxies located in the densest environments compared to the entire sample. However, similar to our findings, the influence of the environment disappears for galaxies with stellar masses exceeding $\simeq10^{10} M_\odot/h $. It is notable that their model also utilizes unstable cold gas to fuel the growth of central black holes, although they do not attribute the observed result to this mechanism. 

\subsection{Convergence mass and SMBH growth}

\label{sec:convergence}

\begin{figure}

\centering
\includegraphics[width=\columnwidth]{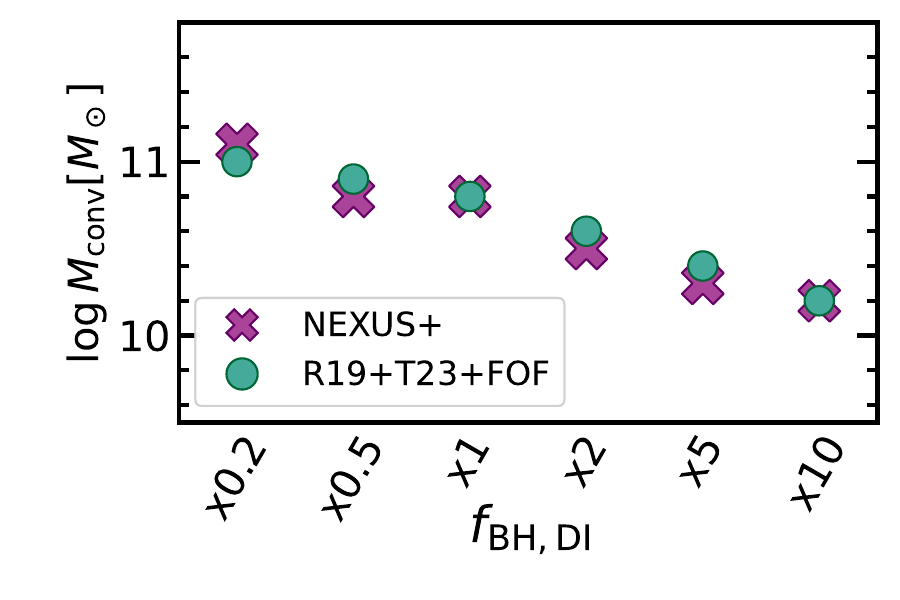}
\caption{Convergence mass for different runs where the parameter ruling the DI-driven SMBH growth ($f_{\rm BH,\, DI}$) is modified by a constant factor, specifically $f_{\rm BH,\, DI}$ x $0.2,\,0.5,\,2,\,5,\,{\rm and}\, 10$. The convergence mass is the stellar mass above which the $M_{\rm stars}-{\rm sSFR}$ relation show no discernible differences among galaxies belonging to different environments, at fixed stellar mass. We report the results obtained with \NEXUS{} environments (\textit{purple crosses}) as well as for the other environments definitions adopted in this work, namely spherical voids and their walls, cylindrical filaments and massive FOF groups (\textit{green circles}, R19+T23+FOF).}
\label{fig:convergence} 
\end{figure}

Based on the previous discussion, it becomes clear that there is a stellar mass threshold beyond which no noticeable distinctions exist among galaxies in various environments, given a constant stellar mass. In our reference model, this \textit{convergence mass} may exhibit slight variations depending on the property under consideration and the method used to identify the different environments. However, it is approximately $M_{\rm conv} \simeq {10^{10.8}}  M_\odot$ for both the sSFR and dust mass (Fig. \ref{fig:scaling_SF_dust}). In Sec. \ref{sec:SMBHgrowth} we have shown that this behaviour is due to the DI-driven SMBH growth channel. This process is regulated by the parameter $f_{\rm BH, \, DI}$, representing the fraction of unstable gas that undergoes accretion onto the BH during episodes of DIs ($f_{\rm BH, \, unst}$ in Eq. 23 \citetalias{Parente2023}). Here we assess the stability of the \textit{convergence mass} to changes in $f_{\rm BH, \, DI}$ when using different environments identifiers: \NEXUS{} on one hand; or specific identifiers for distinct environments $-$ spherical voids, cylindrical filaments and massive halos $-$ on the other hand (R19+T23+FOF).

To this aim, we operationally define the \textit{convergence mass} as follows. We calculate the median trends of the $M_{\rm stars}-{\rm sSFR}$ relation\footnote{A similar approach can be applied to the $M_{\rm stars}-{M_{\rm dust}}$ relation.} for all the environments, as reported in Fig. \ref{fig:scaling_SF_dust}. We employ stellar mass bins of $0.1 \, {\rm dex}$. The \textit{convergence mass} $M_{\rm conv}$ is identified as the smallest $M_{\rm stars}$ value where the median ${\rm sSFR}$ for all environments falls within a range of $0.2\,{\rm dex}$.
We compute $M_{\rm conv}$ for different runs in which we modify only the $f_{\rm BH,\, DI}$ parameter by a constant factor\footnote{\rev{But maintaining its functional dependence on the halo virial velocity.}}, specifically $f_{\rm BH, \, DI}$ x $0.2,\,0.5,\,2,\,5,\,{\rm and}\, 10$. The results for the \textit{convergence mass} obtained in these experiments are reported in Fig. \ref{fig:convergence}.

The general trend suggests that, regardless of the environment classification process, the \textit{convergence mass} decreases as the efficiency of the DI-driven SMBH growth increases. This is because higher values of $f_{\rm BH, \, DI}$ lead to more efficient and rapid SMBH growth, enabling these objects and their associated feedback to be relevant in less massive systems. A departure of a factor of 2 from the fiducial x1 value implies a variation of approximately $0.2\,{\rm dex}$ in $M_{\rm conv}$.\\

It is important to note that the parameters defining $f_{\rm BH, \, DI}$ allow our model to reproduce the local SMF and the fractions of quenched galaxies at different masses \citepalias[Sec. 3.1 of][]{Parente2023} and were not further adjusted in the present work. Interestingly, as will be seen in Sec. \ref{sec:compobs}, the \textit{convergence mass} also manifests in observed galaxy samples. \rev{However, we prefer to use the results in this section as an experiment to improve the understanding of our model's physics rather than using it for direct model refinement.}

\section{Comparison with observations}
\label{sec:compobs}

\begin{table*}
\begin{center}
     \begin{tabular}{c c c c c } 
 \hline
 \hline
 {} & {} & {} & \multicolumn{2}{c}{Number of galaxies} \\
 {Environment} & {Algorithm} & {Criterion} & {SDSS} & {GAMA/\textit{H}-ATLAS} \\
 {} & {} & {} & {catalog} & {sample}\\[0.5ex]
\hline\hline
Voids & \citet{Ruiz2015,Ruiz19} & $r/r_{\rm void} \leq 1$ & 9367 & 95 \\
Walls & \citet{Ruiz2015,Ruiz19} & $1<r/r_{\rm void} \leq 1.2$ & 16863 & 138 \\
Filaments & \cite{Taverna:2023} & $M_{\rm nodes} \geq 10^{13.5}\, M_\odot$& 12556 & 395 \\
Massive groups& \citet{RodriguezMerch20} & $M_{\rm groups} \geq 10^{13.5}\, M_\odot$& 35161 & 380 \\
\midrule[0.2\arrayrulewidth]

\hline\hline
\end{tabular}
\caption{Details on the environment identification performed on the SDSS-DR16 catalog (see Sec. \ref{sec:envclass}). For each environment (voids, walls, filaments, and massive groups), we report the identification method adopted and the number of galaxies identified in the SDSS catalog and GAMA/\textit{H}-ATLAS sample. \rev{Note that the number of galaxies in our simulated sample is much larger (Tab. \ref{tab:N_objs_SAM}).}}
\label{tab:envobs}

\end{center}

\end{table*}

In this section, we compare observations with our SAM predictions. As for our simulated galaxies, in order to highlight the importance of the in situ, DI-driven, SMBH growth channel, we compare the predictions obtained from our fiducial version of the SAM with the aforementioned scenario where SMBH accretion during DIs is switched off ($f_{\rm BH,\, DI}=0$).

Our goal is to compare the environments identified in the simulated and observed samples as fairly as possible, exploiting the identification methods detailed in Sec. \ref{sec:envclass}. Specifically, here we focus on four different environments, ranging from low to high large-scale density:
\begin{itemize}
    \item \textbf{Voids} In both the observed and simulated catalogs we take galaxies residing within the spherical voids as identified by the \cite{Ruiz2015,Ruiz19} algorithm (Sec. \ref{sec:envclass:voids}).
    \item  \textbf{Walls} This category includes galaxies in the proximity $(1 < r/r_{\rm void} \leq 1.2)$ of the aforementioned spherical voids.
    \item \textbf{Filaments} Galaxies within filaments are identified using the \citet{Taverna:2023} algorithm, as applied to both the observed and simulated galaxy samples (Sec. \ref{sec:envclass:filsgroups}).
    \item \textbf{Massive groups} These are galaxies belonging to groups more massive than $M_{\rm group}>10^{13.5}M_\odot$. In the observational catalog the group mass is estimated by the \cite{RodriguezMerch20} algorithm, while in our SAM we take the virial mass of the FOF halo (Sec. \ref{sec:envclass:filsgroups}). We take these galaxies as representative of the densest regions of the large-scale structure.
\end{itemize}

The criteria employed and the number of objects identified in each environment within the observed galaxies catalog are summarized in Table \ref{tab:envobs}.\\
We emphasize that our purpose is not to conduct a one-to-one comparison between environments in observations and simulations. Instead, our main interest is comparing trends among the environments of the two samples.

\subsection{Star formation}
\label{sec:compobs:sf}

\begin{figure*}[htb]
    \centering
    \includegraphics[width=2\columnwidth]{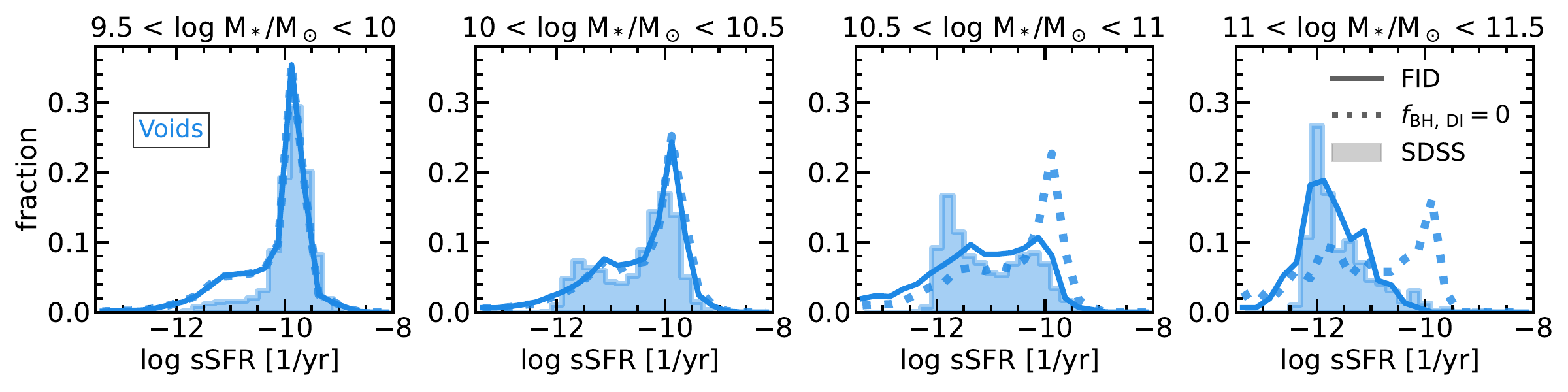}\\
    \includegraphics[width=2\columnwidth]{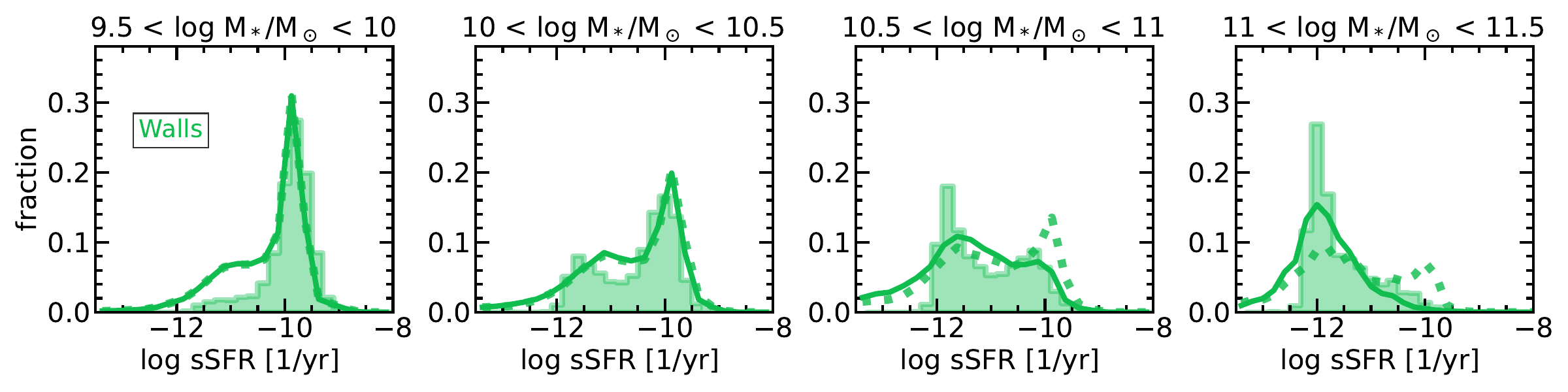}\\
    \includegraphics[width=2\columnwidth]{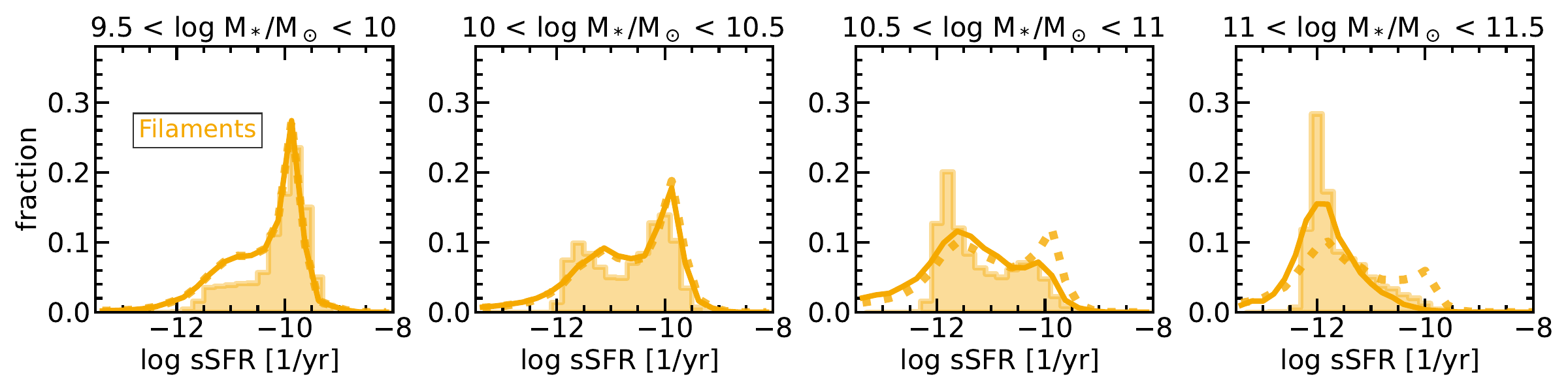}\\
    \includegraphics[width=2\columnwidth]{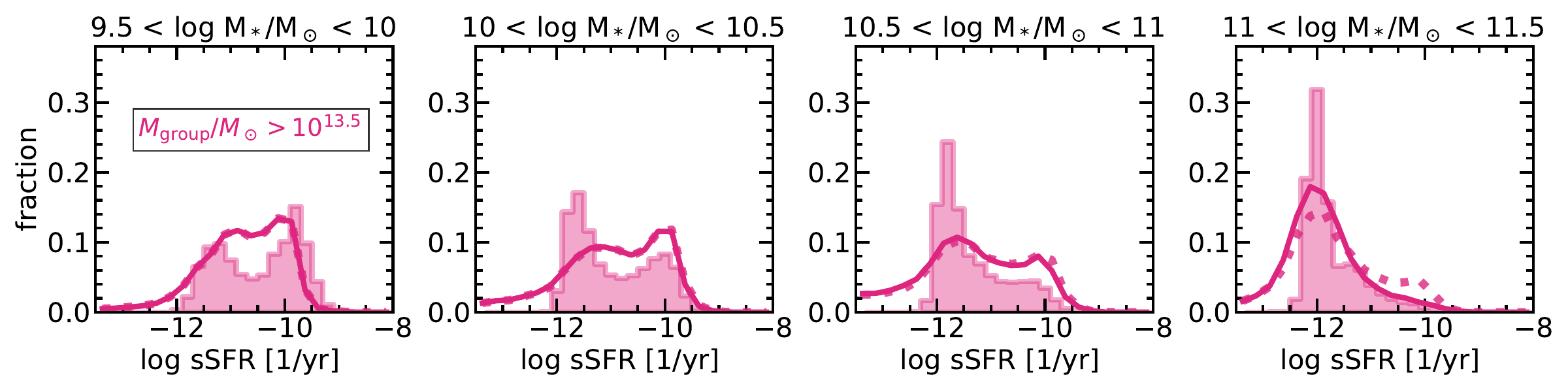}
    \caption{Specific star formation rate distributions for galaxies in different stellar mass bins, namely (from left to right) ${\rm log} \, {M_{\rm stars}/M_\odot} \in (9.5-10),\,(10-10.5),\,(10.5-11),\,(11-11.5)$. Results for different cosmic environments are shown, namely (from top to bottom): voids in blue, walls in green, filaments in orange, and massive groups ($M_{\rm group} > 10^{13.5}\, M_\odot$) in magenta. We report our model predictions obtained with both the fiducial (FID) and \nobhdi version of our SAM (respectively solid and dotted lines). In each panel we compare with SDSS data, shown as filled histograms. Environmental classification in both the observed and simulated catalog is performed as detailed in Sec. \ref{sec:compobs} (see also Table \ref{tab:envobs}). }
    \label{fig:topplot_ssfr}
\end{figure*}

\begin{figure}

\centering
\includegraphics[width=0.99\columnwidth]{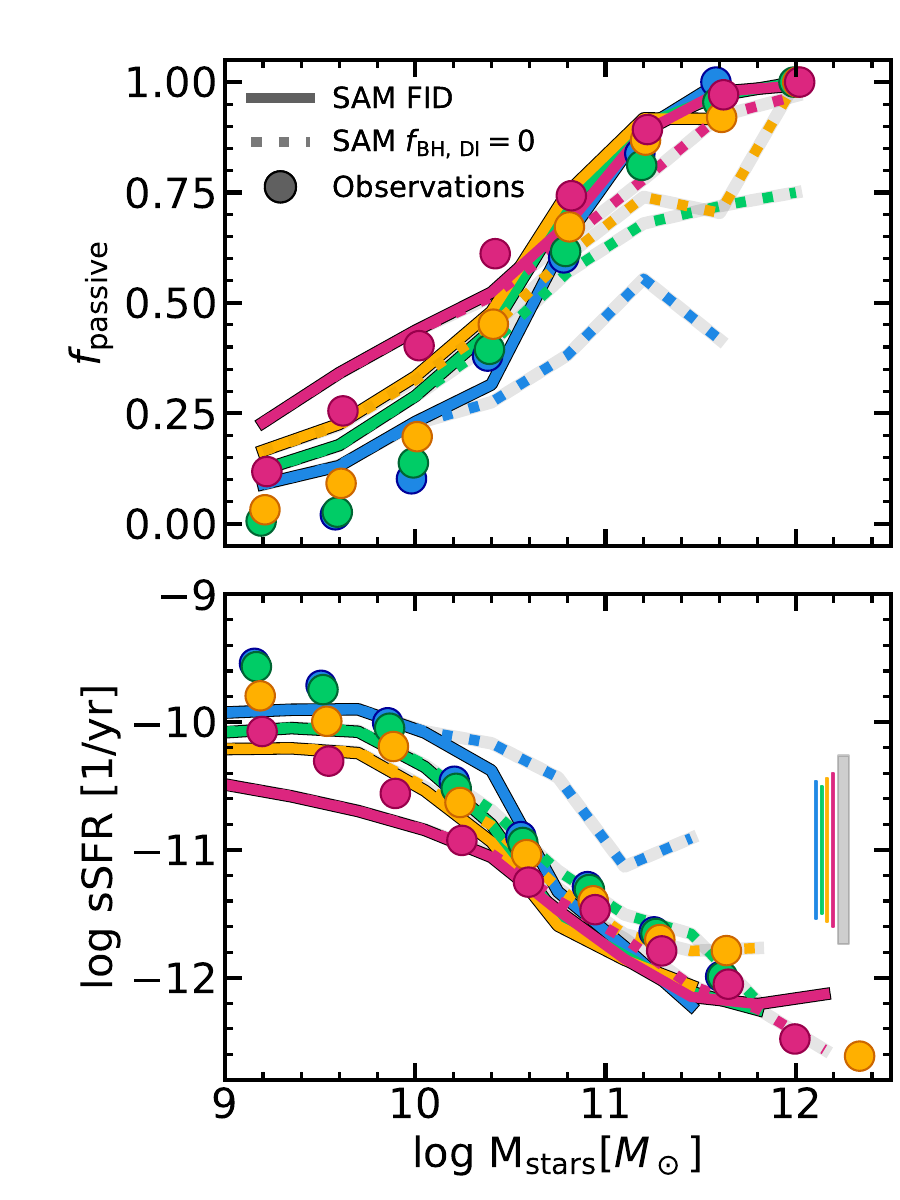}
\caption{Fraction of passive galaxies (${\rm sSFR}<10^{-11}\,{\rm yr}^{-1}$; \textit{top panel}) and sSFR (\textit{bottom panel}) as a function of stellar mass in voids, walls, filaments and massive groups ($M_{\rm group}>10^{13.5}\, M_\odot$). Colors are as in Fig. \ref{fig:topplot_ssfr}. We report our model predictions obtained with both the fiducial (FID) and \nobhdi version of our SAM (respectively solid and dotted lines). Filled circles refer to observations (SDSS data). We also report \rev{the typical $14$-$86$th dispersion of observations (vertical lines) and of our model (grey shaded rectangle)} in the right side of the bottom panel.}
\label{fig:passive-sMS} 
\end{figure}

In Fig. \ref{fig:topplot_ssfr} we show the sSFR distributions for galaxies in different stellar mass bins and cosmic web environments. As widely discussed in literature \citep[e.g.][]{Katsianis21}, this distribution typically features a bimodal behaviour, which corresponds to the superposition of a star-forming (${\rm sSFR} \simeq 10^{-10} \, {\rm yr}^{-1}$) and a passive population (${\rm sSFR} \simeq 10^{-12} \, {\rm yr}^{-1}$) at $z=0$.

From the reported observations it is clear that the larger the stellar mass is, the larger is the prevalence of the passive population \textit{in each environment}. This pattern is even more evident when examining the fraction of passive galaxies $f_{\rm passive}$ (defined as those featuring ${\rm sSFR}<10^{-11}\, {\rm yr}^{-1}$) and the sSFR as a function of stellar mass, both displayed in Fig. \ref{fig:passive-sMS}. The observed $f_{\rm passive}$(sSFR) increases(decreases) with the density of the environment \textit{at fixed stellar mass} for $M_{\rm stars}\lesssim 10^{11}\, M_\odot$.

Our fiducial model displays similar trends, that is an increase of $f_{\rm passive}$ and a decrease of the sSFR with stellar mass and, for $M_{\rm stars}\lesssim 10^{10.8}-10^{11}\, M_\odot$, with the density of the environment. Differences between the FID and \nobhdi outcomes are particularly evident at large $M_{\rm stars}$ and for low density environments. In the two most massive stellar mass bins reported in Fig. \ref{fig:topplot_ssfr}, the \nobhdi  model tends to predict more star forming galaxies than the FID one in both voids and walls. The same holds true when inspecting Fig. \ref{fig:passive-sMS}. Voids galaxies in the \nobhdi model exhibit a too large(small) sSFR($f_{\rm passive}$) with respect to observations at large $M_{\rm stars}$. Instead, the FID model aligns well with SDSS observations in this mass regime. Thus the impact of the SMBH growth during DIs is particularly relevant for star formation quenching in such isolated galaxies.

Also, the passive fraction (and the sSFR, to some extent) of the \nobhdi model exhibits a clear trend with the environment at \textit{any stellar mass}. This behaviour contrasts with observations, where the environmental trend is observed only up to $M_{\rm stars}\lesssim 10^{11} M_\odot$. Conversely, the fiducial model predicts a lack of environmental dependence at large $M_{\rm stars} \, (\gtrsim 10^{10.8}-10^{11} \, M_\odot)$, in keeping with the observed pattern.\\

We conclude that the SMBH growth mechanism adopted by the \nobhdi model is not sufficient to shut down the star formation in massive, isolated galaxies.

\subsection{Dust mass}
\label{sec:compobs:dust}

\begin{figure*}
\centering

\includegraphics[width=0.8\linewidth]{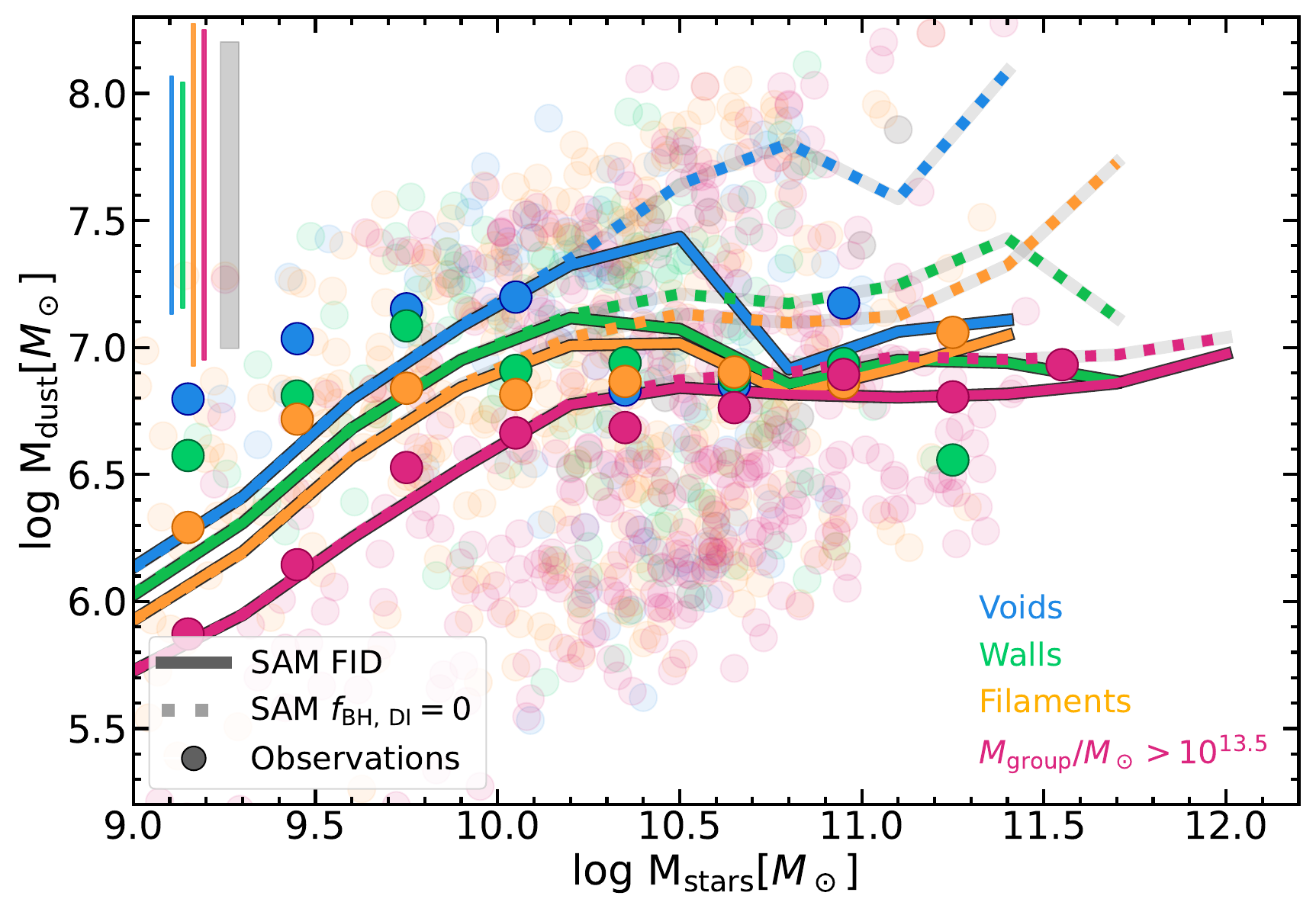}
\caption{Dust $-$ stellar mass relation in different large scale environments: voids in blue, walls in green, filaments in orange, and massive groups ($M_{\rm group} > 10^{13.5}\, M_\odot$) in magenta. The results obtained from our SAM are shown as lines, which refer to medians, while the typical $14$-$86$th dispersion is shown as a grey shaded rectangle in the upper left of the plot. Solid lines refer to our fiducial SAM, while results from the \nobhdi model are shown as dotted lines. 
Observations from the GAMA/\textit{H}-ATLAS sample \citep{Beeston18} are shown as filled points which represent the mean dust mass in stellar mass bins of $0.3\,{\rm dex}$ width, while the whole sample of observations is shown in the background as low opacity circles. We also report \rev{the typical $14$-$86$th dispersion} associated with each environment as vertical lines in the upper left side of the figure. Environments and groups in this figure are identified as detailed at the beginning of Sec. \ref{sec:compobs} (see also Table \ref{tab:envobs}).
}
\label{fig:topplot_dust}
\end{figure*}

Fig. \ref{fig:topplot_dust} compares model predictions on dust masses in local galaxies, as a function of stellar mass and environment, to available observations. The data reported in this section are obtained by matching the SDSS-DR16 catalog and the GAMA/\textit{H}-ATLAS sample, thus the final number of objects is much smaller than the number of SDSS objects shown in the previous section (see Table \ref{tab:envobs}). We focus on the mean trends of data, which feature a large dispersion, keeping in mind the limited sample size.

For $M_{\rm stars} \lesssim 10^{10.5}\, M_\odot$, observations suggest that at fixed stellar mass galaxies are generally more dust rich in less dense environments, with void galaxies featuring a mean dust content $5-10$ times larger than massive groups objects.
The median trends and the environmental differences in this mass range predicted by the fiducial run of our SAM are consistent with observations.\\
At larger stellar masses ($M_{\rm stars} \gtrsim 10^{10.5} \, M_\odot$ in observations and $M_{\rm stars} \gtrsim 10^{10.8} \, M_\odot$ in our model) dust masses appear to be nearly environment-independent both in observations and in the fiducial model. 
By converse, in the \nobhdi model, the objects residing in low density environments feature a far too large median $M_{\rm dust}$. This is consistent with the discussion in Sec. \ref{sec:compobs:sf}, according to which the fraction of highly star forming massive galaxies (likely gas and dust rich) is suppressed in low density environments by the novel in situ SMBH growth.

Although a direct comparison between the \textit{convergence mass} (Sec. \ref{sec:convergence}) in the SAM and observations is not straightforward\footnote{This would need to identify environments in mock catalogues from the simulated galaxy sample in the same way as we do in the observed sample. Also, due to the lower number of objects in the observed catalog the accuracy of its calculation in observations is somewhat unreliable.}, we note that for observed $M_{\rm dust}$ curves, convergence occurs around ${\rm log}{M_{\rm stars}/M_\odot} \sim 10.5-11$, which is in nice agreement with the results obtained from SAM galaxies.\\

\section{Summary and conclusions}
\label{sec:discussionconclusion}

In this work, we study the dependence of local galaxy properties, mainly specific Star Formation Rate (sSFR) and dust, on the large-scale environment, at fixed stellar mass. We adopt the simulated galaxy catalog produced by the semi-analytic model (SAM) introduced in \cite{Parente2023}, which is an extension of the \textsc{L-Galaxies} SAM \citep{Henriques2020}. \rev{This version includes a state of the art model of dust production and evolution in galaxies, as well as an updated treatment of the bulge and SMBH growth during disc instabilities (DIs).} We identify cosmic web environments from the underlying DM simulation exploiting different methods. We employ \NEXUS{} \citep{Cautun2013} as a comprehensive method to dissect the cosmic web into voids, walls, filaments, and nodes. This approach provides us with a unified framework to identify different cosmic environments. In contrast, we also utilize specific algorithms tailored for distinct environments: the method introduced by \cite{Ruiz2015, Ruiz19} for identifying spherical voids and associated walls, the approach outlined in \citet{Taverna:2023} for detecting cylindrical filaments, and the well-known FOF halo finder to pinpoint massive groups ($M_{\rm group}>10^{13.5}\, M_\odot$), representing the most dense environments.\\

First, we study galaxy properties as a function of $M_{\rm stars}$ to highlight the role of the environments (Sec. \ref{sec:galaxyprop}). For galaxies with $M_{\rm stars}\lesssim 10^{10.8}\, M_\odot$ we find a clear and systematic trend: at fixed stellar mass, galaxies in less dense environments feature a larger specific SFR and dust abundance. Contrarily, differences among environments are much less evident in more massive objects. 

We interpret these results in light of the evolution of galaxies with the same $z=0$ stellar mass, but residing in different environments (Sec. \ref{sec:galaxyassembly}). At $M^{z=0}_{\rm stars}\leq 10^{10.5}\, M_\odot$, the less dense is the environment, the slower is the stellar mass evolution. At a given $M^{z=0}_{\rm stars}$, galaxies in less dense environments feature a SFR peak at more recent times, ending up with a larger SFR and dust content. Contrarily, galaxies in more dense environments are in the declining phase of their SF activity. Thus void galaxies host, on average, younger stellar populations, are more star-forming and dust rich with respect to galaxies located in other environments with the same $M^{z=0}_{\rm stars}$. Relevant differences in the stellar mass assembly are not observed instead for more massive galaxies $(M^{z=0}_{\rm stars} \gtrsim 10^{10.8} \, M_\odot)$, and this is in keeping with the environment-insensitivity of star formation and dust for massive objects already discussed. This finding is remarkably robust, irrespective of the variety of methods for environments identification adopted in this work. 

The picture resulting from our model suggests that galaxy properties are affected by the large-scale environment up to a certain \textit{threshold} mass. The key to interpreting this finding stands in the supermassive black hole (SMBH) growth channel during disc instabilities (DIs), which is adopted in our model. This channel enables the growth of SMBHs even in isolated galaxies, allowing them to halt star formation through AGN radio mode feedback. In the absence of this channel, the growth of SMBHs in isolated objects is significantly discouraged, as the other prominent growth channel in our SAM occurs during merger events.
This \textit{in situ} SMBH growth tends to erase differences among galaxies that exceed a specific stellar mass ($M_{\rm stars}\simeq 10^{10.8}\,  M_\odot$, mass above which the SMBH driven quenching becomes relevant in our model) residing in different environments.\\

Finally, we test our results against observations (Sec. \ref{sec:compobs}). We identify different environments in the SDSS-DR16, namely spherical voids, their associated walls, cylindrical filaments, and massive groups $(M_{\rm group}\gtrsim10^{13.5}\, M_\odot)$ using the algorithms by \cite{Ruiz2015}, \cite{Taverna:2023}, and \cite{RodriguezMerch20}. This is conceptually similar to the identification process in our SAM catalog, allowing us to compare the results. The fraction of passive galaxies $({\rm sSFR} < 10^{-11}\, {\rm yr}^{-1})$ with $M_{\rm stars}\lesssim10^{10.5}-10^{11}\, M_\odot$ increases with the increasing density of the environment, in both the SAM and SDSS catalog. As for the most massive objects $(M_{\rm stars} \gtrsim 10^{10.5}-10^{11}\, M_\odot)$, in both our model and observations, we do not find any sign of a prominent star-forming population in any environment. This is a direct consequence of the DI-driven SMBH growth adopted in our model. Without this channel ($f_{\rm BH,\, DI}=0$), our model would produce a prominent star forming population among massive voids galaxies, and this is not present in the observed sample (Figg. \ref{fig:topplot_ssfr} and \ref{fig:passive-sMS}). 

We perform a similar comparison between our model and observations (combining the SDSS-DR16 and GAMA/\textit{H}-ATLAS surveys) in the $M_{\rm stars}-M_{\rm dust}$ plot (Fig. \ref{fig:topplot_dust}). We find a good match, namely the less dense is the environment, the larger is the dust content of galaxies with $M_{\rm stars}\lesssim 10^{10.5} \, M_\odot$ in observations ($M_{\rm stars}\lesssim 10^{10.8} \, M_\odot$ in our model). The $\sim 0.5-1\, {\rm dex}$ difference in the mean dust mass of voids and massive groups galaxies is not observed above this mass. Again, this confirms the importance of the SMBH growth channel in DIs, without which too dust-rich void galaxies would be predicted. \\

We draw two main conclusions from this work. First, we confirm the relevance of the secular process of SMBH growth during DIs in our SAM, since it causes quenching of galaxies in isolated environments. The importance of this growth channel has been pointed out also by recent observational and simulation-based investigations, whose claim is that secular processes dominate over mergers in growing the central SMBH \citep[e.g.][]{Martin18,Smethurst19}. The second conclusion concerns our approach more than our results. Investigating the properties of galaxies in different large-scale environments may be a useful tool for constraining the relevance of in situ and ex situ processes in shaping galaxy evolution. 

The method outlined here can be improved in various ways, for example, by striving for a more accurate match between the environments identified in simulations and observations. A possible way to achieve this may be to identify environments from mock catalogs derived from the SAM (or from hydrodynamical simulations as well) to mimic the identification process performed on the observed catalog as closely as possible. 

In any case, besides any possible improvement, the role of future observational surveys such as Euclid \citep{Laureijs2011}, DESI \citep{desi2016} and LSST \citep{Ivezic2019}, is crucial. They will substantially advance our knowledge of the large-scale structure and, concomitantly, deepen our insight into the evolution of galaxies within distinct cosmic environments.

\begin{acknowledgments}
\small
\rev{We thank the anonymous referee for the constructive report and the suggestions which improved the clarity of our work.\\}
We aknowledge Agustín M. Rodríguez Medrano for sharing the observed catalog of voids, and Angus H. Wright for directing us to the GAMA catalogue. We thank Robert M. Yates for carefully reading the manuscript and providing constructive comments. MP is grateful to Lumen Boco for his stimulating comments and to Meriem Behiri for the help in dealing with observational catalogs.

This project has received funding from the European Union's HORIZON-MSCA-2021-SE-01 Research and Innovation programme under the Marie Sklodowska-Curie grant agreement number 101086388  - Project acronym: LACEGAL, from Consejo Nacional de Investigaciones Cient\'ificas y T\'ecnicas (CONICET) (PIP-2021 11220200102832CO, PIP-2022 11220210100064CO) and from the Ministerio de Ciencia, Tecnología e Innovación (PICT-2020 03690) de la Rep\'ublica Argentina.

Simulations have been carried out at the computing centre of INAF (Italia). We acknowledge the computing centre of INAF-Osservatorio Astronomico di Trieste, under the coordination of the CHIPP project \citep{bertocco2019,Taffoni2020}, for the availability of computing resources and support.
\end{acknowledgments}

%

\vspace{5mm}


\software{astropy \citep{2013A&A...558A..33A,2018AJ....156..123A,Astropy22}, \textsc{TOPCAT} \citep{TOPCAT}, \textsc{L-Galaxies} \citep{Henriques2020}.}



\appendix

\section{Environments Stellar Mass Function and satellites fraction}
\label{app:SMFsat}

\begin{figure}
    \centering

    \includegraphics[width=0.3\columnwidth]{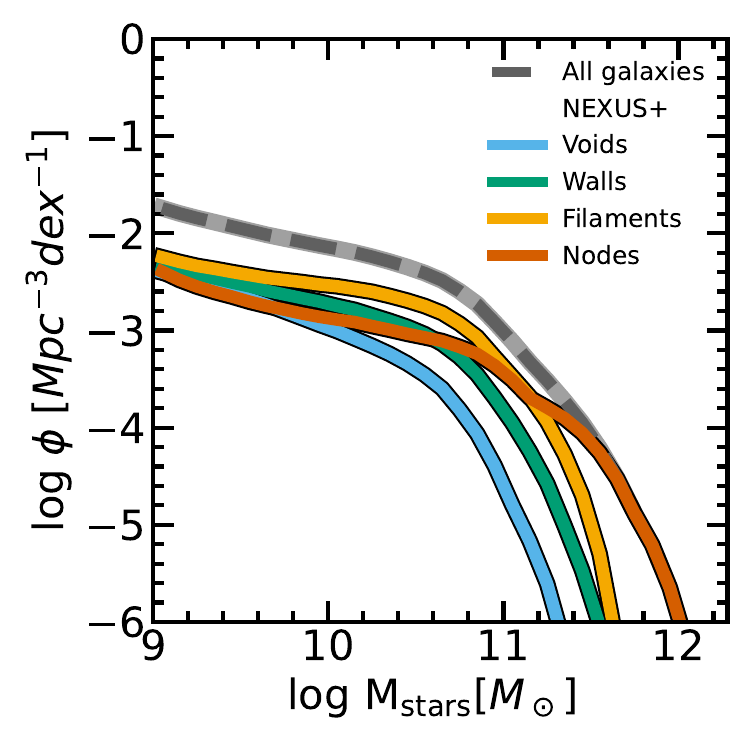} \qquad \qquad    
    \includegraphics[width=0.35\columnwidth]{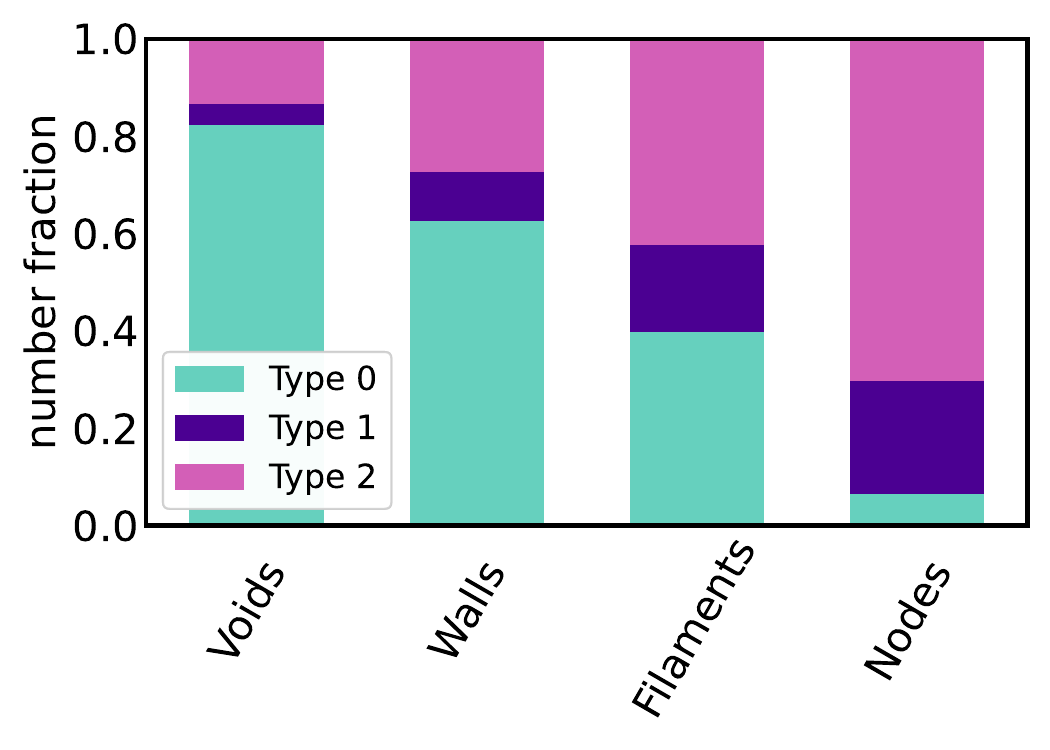}\\
    \includegraphics[width=0.3\columnwidth]{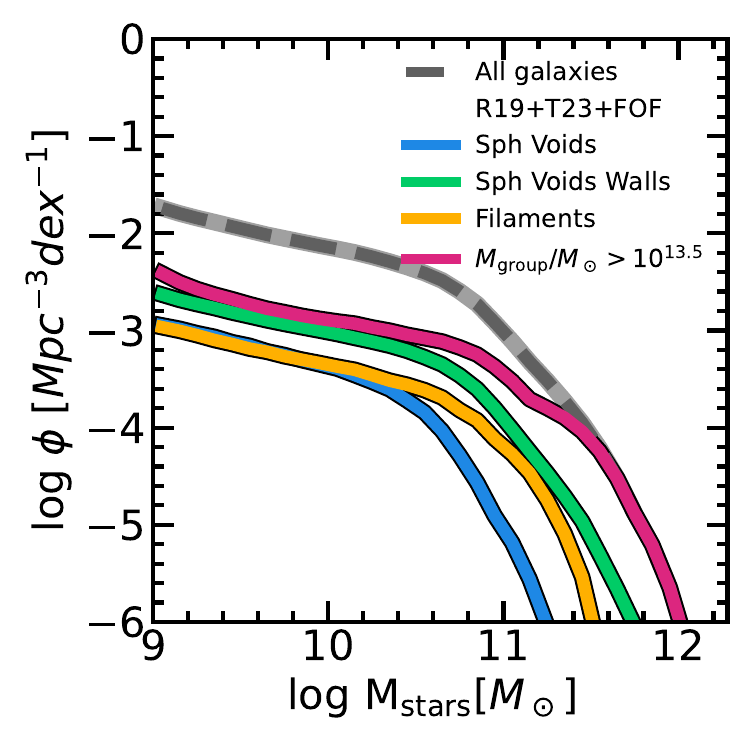} \qquad \qquad    
    \includegraphics[width=0.35\columnwidth]{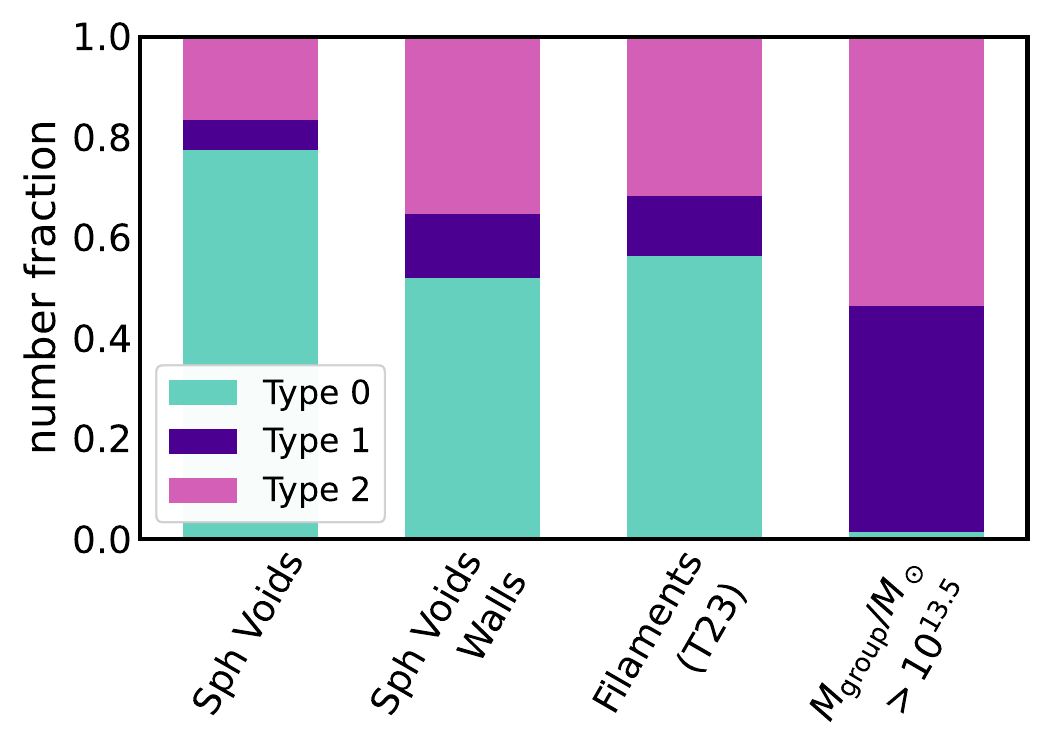}

    \caption{Stellar Mass Function (\textit{left panels}) and number fraction of simulated galaxies of type 0 (centrals, green), type 1 (satellites, purple), and type 2 (orphans, pink) (\textit{right panels}) at $z=0.0$ for different environments. The \textit{top panels} shows the predictions for environments as identified by \NEXUS{} (voids, walls, filaments and nodes), while the \textit{bottom panels} report the other environments definitions adopted in this work, namely spherical voids and their walls, cylindrical filaments and massive groups (R19+T23+FOF). The SMF of the whole sample of galaxies is also shown as a dashed gray line for reference.}
    
    \label{fig:SMF+SatFrac}
    
\end{figure}

\begin{table*}
\begin{center}
     \begin{tabular}{c c c c c c} 
 \hline
 \hline
 {Environment} & {${\rm log}\, {\mathcal{M}}_*/M_\odot$} & & &  {Environment} & {${\rm log} \, {\mathcal{M}}_*/M_\odot$} \\
\hline\hline
Voids & $10.66$ & & & Sph Voids & $10.59$ \\
Walls & $10.84$& & & Sph Voids Walls & $11.02$ \\
Filaments & $10.92$ & & & Filaments (T23) & $11.02$  \\
Nodes & $11.64$ & & & Massive groups & $11.67$ \\
\midrule[0.2\arrayrulewidth]

\hline\hline
\end{tabular}
\caption{Characteristic mass, derived with a Schechter fit, for the SMFs of the different environments discussed in the main text. SMFs are shown in Fig. \ref{fig:SMF+SatFrac}.}
\label{tab:schechter}

\end{center}

\end{table*}

Here we present the Stellar Mass Function (SMF) and the number abundance of type $0,\,1,\, {\rm and}\, 2$ galaxies (respectively centrals, satellites and orphans\footnote{Orphan galaxies are objects that have already lost their DM halo, but still have a baryonic component.} in the \textsc{L-Galaxies} framework) for different cosmic environments at $z=0$. We report the cosmic environments we focus on in the paper: voids, walls, filaments and nodes as identified by \NEXUS{}, as well as spherical voids and their associated walls, cylindrical filaments and massive groups (R19+T23+FOF). Results are shown in Fig. \ref{fig:SMF+SatFrac}. We also perform a fit of the SMFs with a Schechter function \citep{Schechter}:

\begin{equation}
    \phi(m)dm = {\rm ln}(10) \phi^* 10^{(m-m^*)(1+\alpha)} {\rm exp}(-10^{m-m^*}) dm,
\end{equation}

where $m={\rm log}\, M_{\rm stars}/M_\odot$ and $m^*={\rm log}\, \mathcal{M_*}/M_\odot$. The latter is a parameter often referred to as \textit{characteristic mass}. We report this parameter of the SMFs of different environments in Table \ref{tab:schechter}.

As for the \NEXUS{} SMF, the \textit{denser} is the environment, the larger is the number of massive galaxies (the same is observed for the Halo Mass Function; see e.g. Fig. 17 of \citealt{Cautun2013}). The characteristic mass $\mathcal{M_*}$ of the SMF as well increases with the density of the environment (Tab. \ref{tab:schechter}). The Nodes SMF exhibits a feature at log $M_{\rm stars}/M_\odot \simeq 11.5$. This is due to one of the parameters required by \NEXUS{}, that is the minimum mass for a cell to be classified as a node ($M_{\rm vir} \geq 2\cdot 10^{14}\,M_\odot$ here, about $M_{\rm stars}\gtrsim 2.5 \cdot 10^{11}\, M_\odot$). For this reason, every cell containing objects which meet this criterion will automatically be classified as a node.

Also, the relative fraction of central, satellites and orphan galaxies is sensitive to the environment (top right panel of Fig. \ref{fig:SMF+SatFrac}). The denser is the environment, the smaller is the number of central (type $0$) galaxies. More dense environments, particularly nodes, typically host more groups and clusters of galaxies, and this explains the progressively larger number of satellite (and orphan) galaxies.

A similar pattern is observed when comparing R19+T23+FOF environments (bottom panels). We note that in this case walls galaxies are typically more massive than filaments, and also host a lower fraction of central (type 0) galaxies.

\section{Dust properties across environments}

\label{sec:dustprop}


\begin{figure*}
    \centering

    \includegraphics[width=0.3\columnwidth]{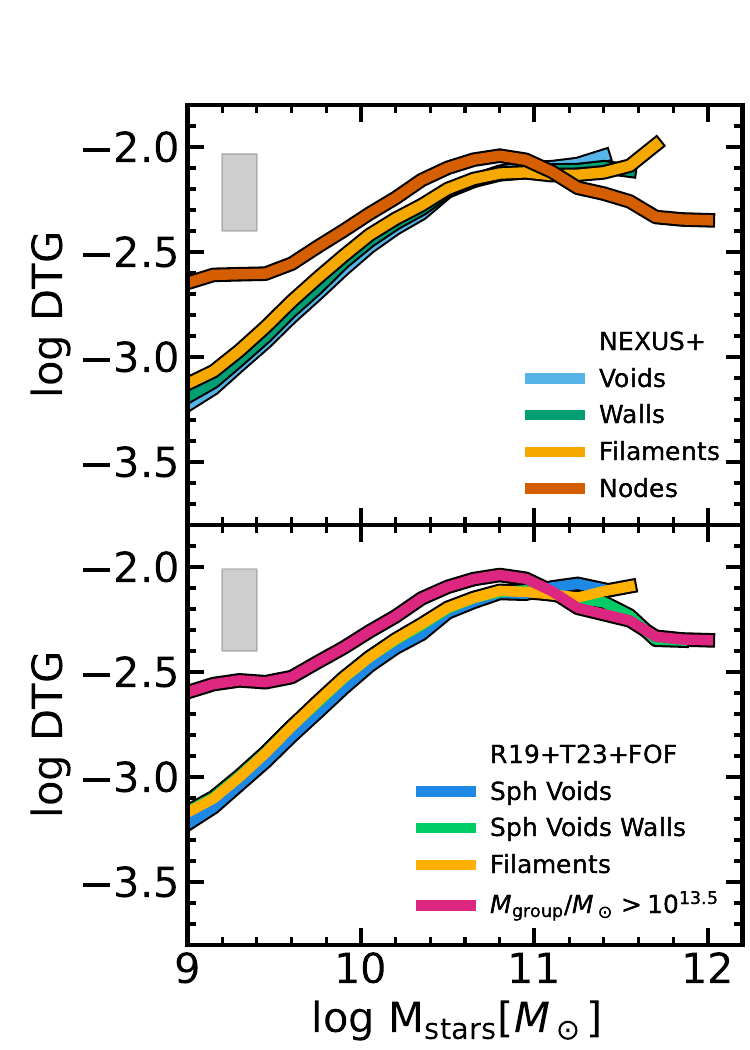} 
    \includegraphics[width=0.3\columnwidth]{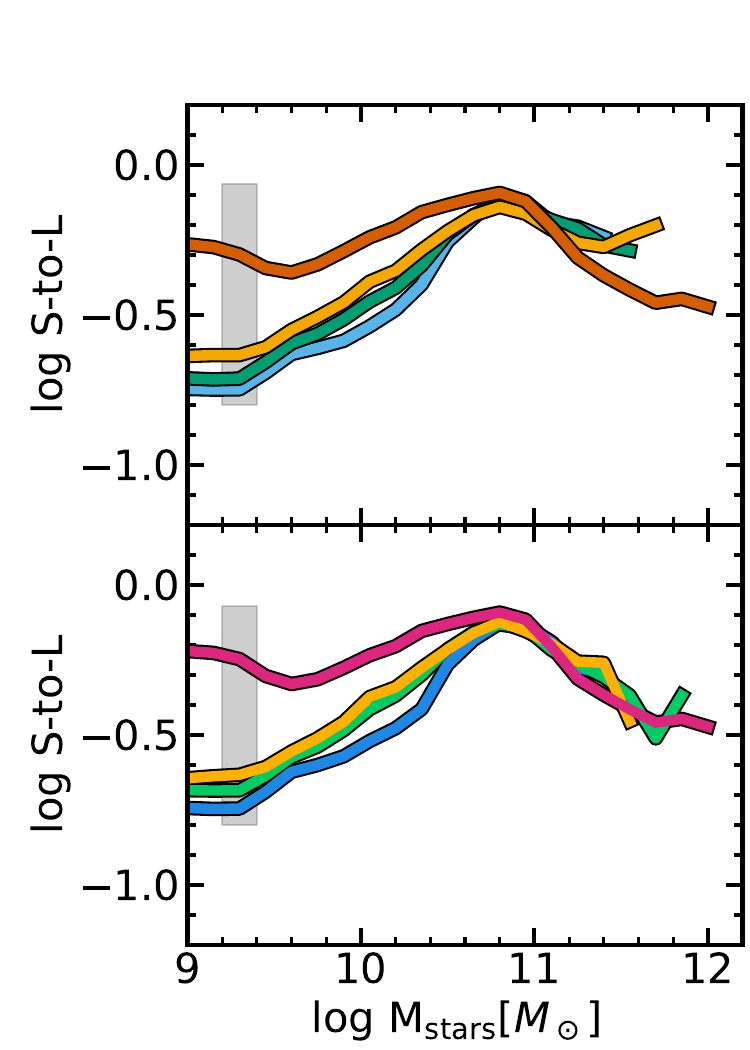}
    \includegraphics[width=0.3\columnwidth]{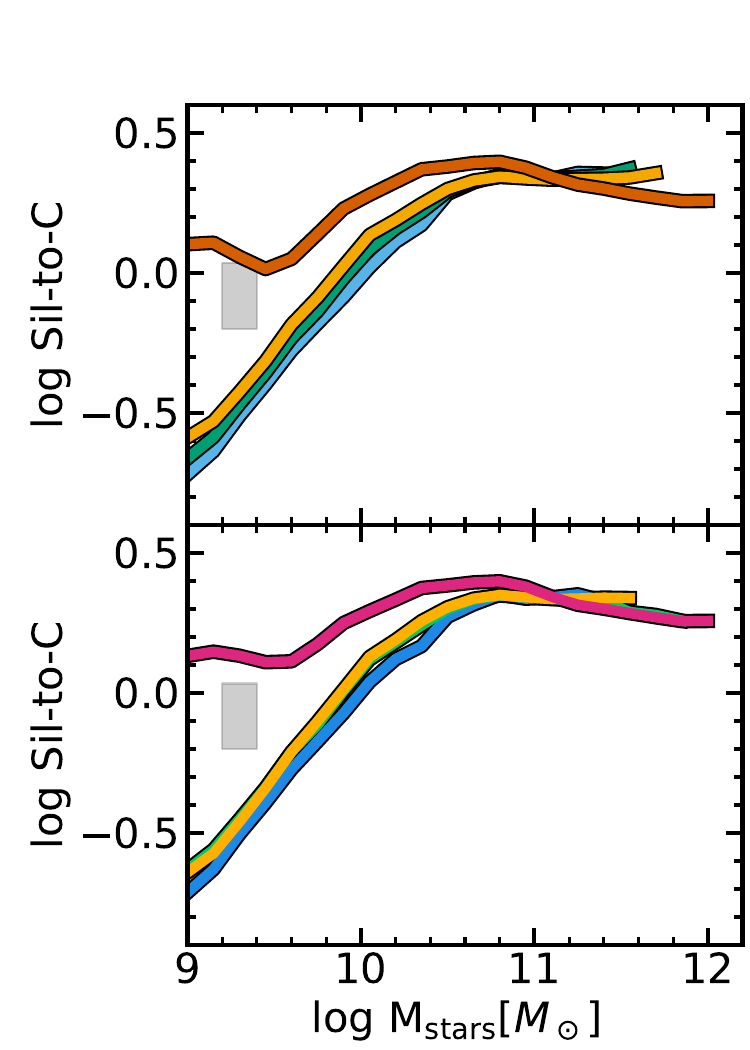}
 
    \caption{Dust-to-Gas (\textit{left panel}), Small-to-Large (\textit{middle panel}), and Silicate-to-Carbonaceous (\textit{right panel}) ratio as a function of stellar mass, in different cosmic web environments. We report the results for \NEXUS{} environments (voids, walls, filaments and nodes, \textit{top panels)}, as well as for the other environments definitions adopted in this work, namely spherical voids and their walls, cylindrical filaments and massive groups (R19+T23+FOF, \textit{bottom panels}).}
    \label{fig:scaling_dust}
\end{figure*}

Here we analyze the dusty properties of galaxies in different environments. Namely, we look at the Dust-to-Gas (DTG), Small-to-Large (S-to-L) and Silicate-to-Carbonaceous (Sil-to-C) ratios as a function of stellar mass. These quantities are particularly relevant as strongly dependent on the grains accretion process in galaxies \citep[e.g.][]{Hirashita11,Aoyama17,Parente2022,Yates2023}. In particular, in our model the accretion process boosts the DTG ratio, as well as the silicate and small grains abundance (the latter up to a certain mass when coagulation becomes more efficient; see e.g. also \citealt{Hou2019}). 

Results are shown in Fig. \ref{fig:scaling_dust}. Although to a lesser extent, the environment still has an impact on these properties: at fixed $M_{\rm stars}$, the denser is the environment, the larger are the DTG, S-to-L and Sil-to-C ratios at log $M_{\rm stars}/M_\odot \lesssim 10.8$. This is particularly evident for galaxies in nodes, which is the curve that deviates the most from the others. However, this behaviour is mainly due to the contribution of satellite galaxies, which are more abundant in dense environments, especially orphan galaxies (see Fig. \ref{fig:SMF+SatFrac}). Indeed, when looking at the same relations for central galaxies only (type 0), differences among environments are strongly suppressed (not shown here). 

Thus the enhancement of DTG ratio at low masses (as well as of S-to-L and Sil-to-C) towards denser environments is strictly related to the treatment of satellites within the SAM. These objects are subject to hot gas stripping. Moreover, orphan galaxies (type 2) have no hot gas at all. As a result, in satellite galaxies, hot gas cooling is strongly (often totally) suppressed, and consequently the cold gas metallicity is not diluted by the inflow of (almost) pristine material. Thus the larger cold gas metallicity of these objects enhances the grains accretion, and consequently DTG ratio, silicate and small grains abundance. 

At ${\rm log}\,M_{\rm stars}/M_\odot\gtrsim 10.8$ trends reverse, with galaxies in denser environments featuring lower values of DTG, S-to-L and Sil-to-C. The large majority of galaxies in this mass range are central galaxies (type 0), which are quenched by the SMBH radio-mode feedback. Consequently, a significant number of galaxies experience a substantial decrease in their molecular gas fraction, inhibiting the grains accretion process and the conversion of gas-phase metals into dust grains (a process that would lead to a DTG increase). Importantly, the prevalence of galaxies with low molecular gas fraction is linked to the environment, with approximately $\sim 95\%$ of galaxies in Nodes (and $\sim40\%$ in Voids) featuring $f_{\rm H_2} \lesssim 5\%$ for $M_{\rm stars} \gtrsim 10^{11} \, M_\odot$. Indeed, galaxies in denser environments typically start to be quenched at earlier times, ending up to $z=0$ with a lower gas, and H$_2$, content. We observe this reversal of trends also in the S-to-L and Sil-to-C trends, which are influenced as well by the accretion process.




\bibliography{SAM+env}{}
\bibliographystyle{mnras}



\end{document}